\documentclass[pre,twocolumn,superscriptaddress]{revtex4}

\usepackage{amsmath,amssymb}
\usepackage[english]{babel} 
\usepackage{graphicx,color}
\usepackage{amsmath} 
\usepackage[T1]{fontenc}

\usepackage{psfrag}

\begin{document}

\title{Stochastic resetting by a random amplitude}

\author{Marcus Dahlenburg}
\affiliation{University of Potsdam, Institute for Physics \& Astronomy,
14476 Potsdam, Germany}
\affiliation{BCAM-Basque Center for Applied Mathematics,
48009 Bilbao, Basque Country, Spain}
\author{Aleksei V. Chechkin}
\affiliation{University of Potsdam, Institute for Physics \& Astronomy,
14476 Potsdam, Germany}
\affiliation{Akhiezer Institute for Theoretical Physics, 61108 Kharkov, Ukraine}
\author{Rina Schumer}
\affiliation{Desert Research Institute, Reno, NV 89512, USA}
\author{Ralf Metzler}
\email{rmetzler@uni-potsdam.de}
\affiliation{University of Potsdam, Institute for Physics \& Astronomy,
14476 Potsdam, Germany}

\date{\today}

\begin{abstract}
Stochastic resetting, a diffusive process whose amplitude is "reset" to the
origin at random times, is a vividly studied strategy to optimize encounter
dynamics, e.g., in chemical reactions. We here generalize the resetting step
by introducing a random resetting amplitude, such that the diffusing particle
may be only partially reset towards the trajectory origin, or even overshoot
the origin in a resetting step. We introduce different scenarios for the
random-amplitude stochastic resetting process and discuss the resulting
dynamics. Direct applications are geophysical layering (stratigraphy) as well
as population dynamics or financial markets, as well as generic search
processes.
\end{abstract}

\maketitle

\section{Introduction}

Albert Einstein \cite{einstein} established the probabilistic approach to
Brownian motion based on the assumption that individual displacements of the
tracer particle are independent (uncorrelated) beyond a microscopic correlation
time, identically distributed, and characterized by a finite variance. This
"schematisation \ldots represents well the properties of real Brownian motion"
\cite{levy}. The theoretical description of stochastic processes, based on the
formulation of fluctuating forces by Paul Langevin \cite{langevin}, is by now
one of the cornerstones of non-eqiulibrium physics \cite{brenig,zwanzig,bennaim},
with a wide field of applications across the sciences, engineering, and beyond.

An important application of diffusive dynamics is in the theory of search processes
\cite{olivier}.
Random search strategies are efficient processes when prior information about
the target is lacking \cite{info,opti} or when the searcher itself can only
move diffusively, such as molecular reactants \cite{biochem}. A number of
specific strategies have been studied as generalization of the classical
Brownian search \cite{smoluchowski}, such as L{\'e}vy flights \cite{lf,lf1},
intermittent search \cite{int,int1}, or facilitated diffusion \cite{bvh,mich}.
Applications of these strategies are found in biochemistry \cite{biochem,otto},
biology \cite{bio}, computer science \cite{comp}, or economy \cite{eco}.

Effects of "resetting" events, when a stochastic process is returned to its
original state, were studied in a neuron model \cite{old_reset_1} and in the
context of multiplicative processes \cite{old_reset_2}. In the seminal work by
Evans and Majumdar \cite{Evans1} "stochastic resetting" (SR) was defined as the
stochastic interruption of a random motion, resetting the particle to its initial
position and starting the process anew. A particular feature is that the mean
first passage time in diffusive search becomes finite and can be minimized
\cite{Evans2}. SR is thus widely applied to search processes.

SR has two random input variables. One is the particle's random motion between
resets, for which numerous processes were considered \cite{DP_1,DP_2,DP_3,DP_4,
DP_5,DP_6,DP_7,DP_8,DP_9}. The other variable describes the stochastic time
span between successive resets, with a variety of studied distributions
\cite{RIL_1,RIL_2,RIL_3,RIL_4,RIL_5, RIL_6,RIL_7}. Concrete SR mechanisms include
resetting to an initial distribution \cite{Evans2} to the previous maximum
\cite{RM_prevmax}, resetting with a memory \cite{RM_memory}, resetting after a
delay \cite{DP_2,RM_del1,RM_del2,RM_del3,RM_del4}, space-time coupled resets
\cite{DP_7,DP_8,RM_stcouple1, RM_stcouple2,non-local,non-local1}, and
non-instantaneous resetting.
SR in confinement was considered for different
dimensions \cite{SC_1}, with different boundary conditions \cite{DP_3,SC_2,SC_3},
or in a potential \cite{SC_4,SC_5,SC_6,SC_7}. Finally interacting particle effects
were studied \cite{I_1,I_2,I_3,I_4}. Applications of SR were discussed in the
context of web search in computer science \cite{App_1,App_2}, enzymatic velocity
\cite{RM_del1,App_3}, reaction-diffusion processes with stochastic decay
\cite{App_4}, backtrack recovery by RNA polymerase \cite{App_5}, and pollination
strategies \cite{App_6}. The first experimental realization of SR was achieved
by tracing diffusing colloidal particles reset by switching holographic optical
tweezers \cite{App_exp}.

Here we consider a random-amplitude SR (RASR), motivated by geophysical
stratigraphic records \cite{HE,SD_classic1}, made up of the layers of sedimentary
material that accumulated in depositional environments but were not subjected
to subsequent erosion. These layers ("beds") are separated by erosional
surfaces where previously existing material was removed by chemical reaction or
physical forces. The periods of time missing from the geologic record due to
erosion are known as "stratigraphic hiatuses" \cite{SD_classic2}. It was in fact
Hans Einstein, Albert Einstein's son, who applied probabilistic approaches to
stratigraphic records \cite{HE}. Geologists
use the stratigraphic record to infer earth's history, and sediment bed type is
used to interpret the depositional setting (river, delta, lake, dune, etc.).
If sediment at multiple points within the stratigraphic column can be dated
using geochronological techniques such as C14 dating \cite{SD_C14}, average
linear rates of accumulation can be calculated. These rates may be serve as
proxies for external forcing such as climate regime.

The generation of the stratigraphic record is typically modeled as a random
process. Thus, random surface elevation at a given point on the earth moves
upward (by deposition), stays constant (no erosion or deposition), or decreases
(erosion). Deposition and erosion are continuous and were described by different
stochastic processes, starting with the work of Kolmogorov \cite{SD_1}. Since
then a variety of stochastic models (i.a., random walks \cite{SD_2} or fractional
Brownian motion \cite{SD_3}) were used to probe the fidelity of the stratigraphic
record with respect to earth history. The observation that measured linear rates
of accumulation decrease as a power-law with measurement interval in a variety of
geologic settings \cite{SD_4}, was attributed to power-law hiatus lengths, which
in turn arise because they are created by return times of random surface
fluctuations \cite{SD_5}. Here we explore an additional mechanism for erosion,
typical for regular (e.g., seasonal) or irregular massive erosion events, such as
extreme rainfall, storms, or floods. In these cases the surface is eroded away by
a sizeable amount during a short period in time. The exact erosion height will be
different each time. We model such extreme events by RASR: resetting occurs at
random intervals with random amplitude (Fig.~\ref{paper_trials_dep_indep}). The
guiding example we consider in the following is that of ballistic propagation of
the process, interrupted by RASR events. Such ballistic motion
may reflect ongoing accretion, for instance due to deposits in a riverbed or
a river delta. Occasional extreme rainfalls or snowmelts cause significant
erosion of these layers, corresponding to the resetting events.

\begin{figure}
\includegraphics[height=8.8cm,angle=270]{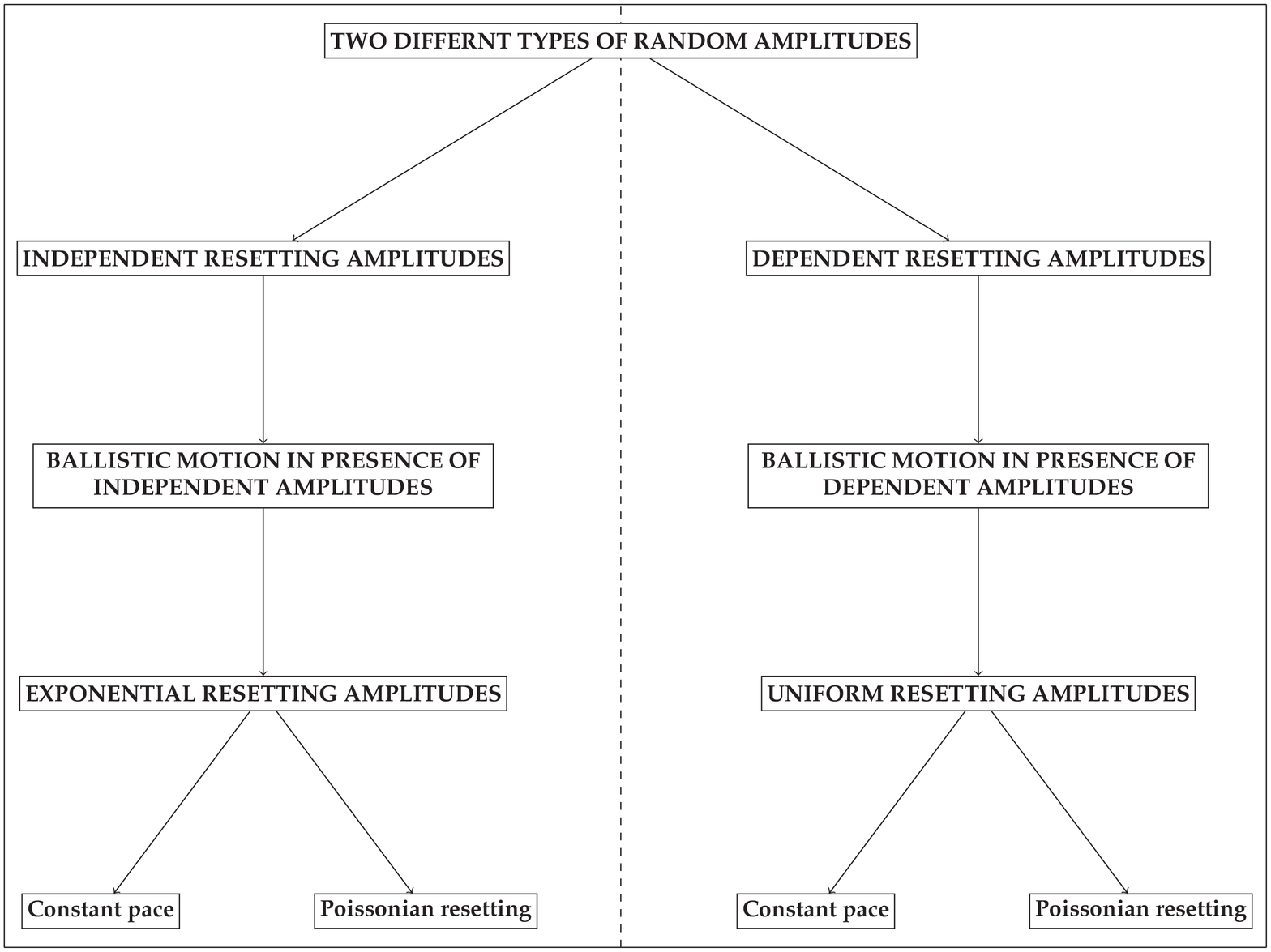}
\caption{Flowchart of the two main concepts, independent and dependent random
amplitude stochastic resetting (RASR) with specific choices of the resetting
and propagation statistics.}
\label{scheme}
\end{figure}

We here develop the RASR model and discuss a range of applications going beyond
the geophysical erosion picture drawn here. Examples include the dynamics of
financial markets hit by occasional crises \cite{AGSR_1,AGSR_2}, population
dynamics affected by partial extinction \cite{AGSR_3}, or germs affected by
antibiotic treatment \cite{AGSR_4}. We note that we call RASR a resetting
process despite the fact that the "reset" leads to a random position.
However, the RASR process keeps the idea of classical resetting in that the
propagation of the test particle is occasionally interrupted by a significant
shift. In the search context mentioned above the RASR process thus represents
a new class of intermittent search processes in which the searcher does not
intermittently return to its "nest" but restarts its search at a range of
key points (points of previous search success, etc.).

The layout of the paper is as follows (compare also the scheme in figure
\ref{scheme}). We first develop the general resetting picutre of our RASR
model in Section \ref{general}. Section \ref{indep} introduces the concept
of indepentent resetting, in which the coordinate of the process does not
depent on the position before resetting. The opposite case, dependent
resetting is developed in Section \ref{dep}. In both cases we consider
specific cases for the timing of the resets and the resetting amplitude
statistic. We draw our Conclusions in Section \ref{conc}, some additional
derivations are deferred to the Appendices.

\section{General resetting picture}
\label{general}

\begin{figure}
\includegraphics[width=8.8cm]{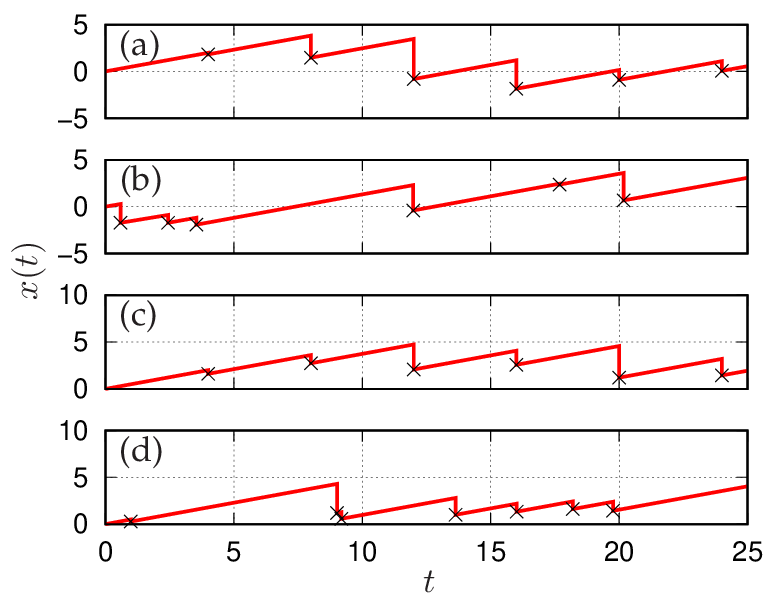}
\caption{RASR sample paths with ballistic displacement ($v=0.5$) and independent
(Poissonian with mean $\zeta=1.6$, panels (a) and (b)) as well as dependent
(uniformly distributed, panels (c) and (d)) resetting amplitudes. In (a) and (c)
resetting events ($\times$) occur at constant pace, in (b) and (d) with
Poissonian waiting times, both with mean rate $r=1/4$.}
\label{paper_trials_dep_indep}
\end{figure}

In the \emph{RASR model\/} $\psi(t)$ denotes the probability density function
(PDF) of time spans between resetting events, and the PDF for the time $t$ at
which the $n$th resetting event occurs is
\begin{equation}
\label{psi_n}
\psi_n(t)=\int_0^t\psi_{n-1}\left(t-t'\right)\psi(t')dt'
\end{equation}
with $\psi_0(t)=\delta(t)$. In Laplace space, therefore, $\tilde{\psi}_n(s)=\tilde{
\psi}^n(s)$. The probability
\begin{equation}
\label{Psi}
\Psi(t)=1-\int_0^t\psi(t')dt'
\end{equation}
of no reset up to $t$
becomes $\tilde{\Psi}(s)=(1-\tilde{\psi}(s))/s$. Finally, the probability to have
exactly $n$ resets up to $t$ is
\begin{equation}
\label{Phi_n}
\Phi_n(t)=\int_0^t\psi_n(t')\Psi(t-t')dt'.
\end{equation}
In what follows we consider independent, identically distributed (iid) resetting
time intervals by using the examples of constant interval lengths ("constant pace")
and Poisson-distributed intervals. The RASR process can have independent resetting
amplitudes $z_n$ at the $n$th step that do not have a lower bound
(Fig.~\ref{paper_trials_dep_indep}a, b). For dependent (bounded) resetting
amplitudes the process never crosses to negative heights $x(t_n)$ 
(Fig.~\ref{paper_trials_dep_indep}c, d).

Let the term $x(t)|x(t_0)$ denote the position $x$ at a certain
time $t$ provided that at time $t_0$ the position was $x_0=x(t_0)$.
For the derivations of the "first resetting picture" we will use the general
relation \footnote{In the classical resetting framework, in which the particle
is returned to its initial position each time, this approach is reduced to the
"first renewal picture" defined in Ref.~\cite{RIL_3}. The same holds for the
term "last resetting picture" introduced below.}
\begin{eqnarray}
\label{condition_first_app}
x(t)|x(t_0)=\left\{\begin{array}{ll}
y(t)|x(t_0)&\mbox{with probability } \Psi(t-t_0)\\
&\mbox{ for } t_0\le t,\\[0.32cm]
x(t)|x(t_1)&\mbox{with probability}\\
&\int\limits_{t_0}^t dt_1 \psi(t_1-t_0)
\end{array}
\right.\
\end{eqnarray}
Eq.~(\ref{condition_first_app}) shows two possibilities. The upper line
describes the possibility of no reset in $[t_0,t]$ with the corresponding
probability $\Psi(t-t_0)$. In this scenario the process, starting at position
$x_0=x(t_0)$ at time $t_0$, fulfils a specific displacement process $y(t)$. Thus,
with probability $\Psi(t-t_0)$ the process $x(t)=y(t)$, which is stochastically
described by $G(y,t;x_0,t_0)$. The lower line of Eq.~(\ref{condition_first_app})
describes the first resetting point
$x(t_1)$ at the random resetting event $t_1$ as a new initial condition of $x(t)$.
The new initial condition $x_1$ at $t_1$ will be described by the distribution
$\phi(x_1,t_1;x_0,t_0)$, which is, without loss of generality, \emph{dependent\/}
on the previous initial condition $x_0$ at $t_0$. The corresponding probability
for this event is $\int_{t_0}^tdt_1\psi(t_1-t_0)$ for $t_1\in[t_0,t]$.

With Eq.~\eqref{condition_first_app} we can find the expression for the
corresponding PDF $P(x,t;x_0,t_0)$,
\begin{eqnarray}
\nonumber
P(x,t;x_0,t_0)&=&\Psi(t-t_0)G(x,t;x_0,t_0)\\
\nonumber
&&\hspace*{-2.2cm}+\int_{t_0}^t dt_1\psi(t_1-t_0)\int_{-\infty}^\infty dx_1
\phi_1(x_1,t_1;x_0,t_0)\\
&&\hspace*{-1.6cm}\times P(x,t;x_1, t_1).
\label{p_r_general_app}
\end{eqnarray}
In Eq.~\eqref{p_r_general_app}, $\phi_1(x_1,t_1;x_0,t_0)$ is the distribution
of the first resetting point $x_1=x(t_1)$ at time $t_1$ under the condition that
the process started at position $x_0$ at time $t_0$. The computation of $\phi_1
(x_1,t_1;x_0,t_0)$ depends on which kind of resetting mechanism we will use.

\section{Independent resetting picture}
\label{indep}

For independent resetting the height after the $n+1$st resetting event is
\begin{equation}
\label{indep_iter_app}
x(t_{n+1})=y(t_{n+1})|x(t_n)+z_{n+1}
\end{equation}
with the initial condition $x(t_0)=x_0$. Here $y(t_{n+1})|x(t_n)$ defines the
unperturbed motion during the time interval $t_{n+1}-t_n$ starting from point
$x(t_n)$. Moreover, $z_{n+1}$ is an iid resetting amplitude of negative value,
$z_i\in(-\infty,0)$. This setup corresponds to our picture of sudden massive
erosion, population decimation, or financial market loss, in which the resetting
amplitude is viewed independent of the process. Conceptually, this type of RASR
corresponds to jump diffusion with one-sided jump lengths \cite{Kou,porporato}.

For $n=0$, Eq.~\eqref{indep_iter_app} yields
\begin{equation}
\label{indep_iter_1_app}
x(t_1)=y(t_1)|x_0+z_1.
\end{equation}
The sum of two random variables implies the convolution of the corresponding
PDFs. Thus, with Eq.~\eqref{indep_iter_1_app}, $\phi_1(x_1,t_1;x_0,t_0)$ is
\begin{equation}
\label{dist_x_1_indep}
\phi_1(x_1,t_1;x_0,t_0)=\int_{-\infty}^\infty dyG(y,t_1;x_0,t_0)q(x_1-y).
\end{equation} 
The PDF $P(x,t;x_0,t_0)$ to propagate from $x_0$ at $t_0$ to $x(t)$ is obtained
by plugging relation (\ref{dist_x_1_indep}) into Eq.~(\ref{p_r_general_app}),
yielding
\begin{eqnarray}
\nonumber
P(x,t;x_0,t_0)&=&\Psi(t-t_0)G(x,t;x_0,t_0)\\
\nonumber
&&\hspace{-1.8cm}+\int_{t_0}^tdt_1\psi(t_1-t_0)\int_{-\infty}^{\infty}dy
G(y,t_1;x_0,t_0)\\
&&\hspace*{-1.2cm}
\times\int_{-\infty}^{\infty}dx_1q(x_1-y)P(x,t;x_1,t_1).
\label{first}
\end{eqnarray}
The first term on the right involves the PDF $G(x,t;x_0,t_0)$ for undisturbed
motion without resetting, where the probability $\Psi(t)$ denotes no resetting during
the time from $t_0$ to $t$. The second term describes free propagation from
$(x_0,t_0)$ to the first resetting point at $(x_1,t_1)$, at which a reset to
$x_1$ occurs with the amplitude PDF $q(x_1-y)$. Then the process is propagated
by $P(x,t;x_1,t_1)$. Eq.~(\ref{first}) can be iterated to include all resetting
steps. From that derivation one can see that the PDF
$P(x,t;x_0,t_0)$ is homogeneous, $P(x,t;x_0,t_0)=P(x-x_0,t-t_0;0,0)$, exactly
when $G$ is homogeneous. In the setting of Eq.~(\ref{first}) we can
describe a general resetting process with arbitrary propagation and independent
resetting events. The first resetting picture described here can be shown to be
identical to the "last resetting picture", as demonstrated for independent
resetting in Apps.~\ref{identity_equal} and \ref{indep_equal}. We now consider
special cases for the propagation, resetting times, and amplitudes.

\subsection{Ballistic propagation}

An illustrative example is given by ballistic propagation (and in fact a special
case of the jump process considered in \cite{porporato}) with speed $v$,
$G(x,t)=\delta(x-vt)$, where we set $x_0=0$ and $t_0=0$. To compute the
characteristic function ${\hat P}(k,t)=\int_{-\infty}^{\infty}dx\exp(ikx)
P(x,t)$ of $P(x,t)=P(x,t;x_0=0,t_0=0)$ for the \emph{first resetting 
picture} (\ref{p_r_general_app}) respectively the \emph{last resetting
picture\/} \eqref{p_r_indep_last_app} in presence of a ballistic propagation,
we use Eq.~(\ref{p_r_general_app}) with $G(x,t;y,\tau)=\delta\left(x-y-v\left(
t-\tau\right)\right)$. The Laplace transform $\tilde{\hat P}(k,s)=\int_{0}^{
\infty}dt\exp(-st) {\hat P}(k,t)$ of the characteristic function ${\hat P}(k,t)$
then reads
\begin{eqnarray}
\nonumber
P\left(x,t\right)&=&\Psi(t)\delta(x-vt)+\int\limits_0^tdt_1\psi (t_1)\int\limits
_{-\infty}^\infty dy\delta(y-vt_1)\\
\nonumber
&&\times\int\limits_{-\infty}^\infty dx_1q( x_1-y) P( x-x_1, t-t_1),
\end{eqnarray}
from which we obtain the Fourier transform
\begin{eqnarray}
\nonumber
{\hat P}(k,t)&=&\Psi(t)\exp(ikvt)\\
&&\hspace*{-1.2cm}+\int\limits_0^tdt_1\psi(t_1)\exp(ikvt_1)\hat{q}(k)\hat{P}
(k,t-t_1).
\end{eqnarray}
Finally, after an additional Laplace transform,
\begin{equation}
\tilde{{\hat P}}(k,s)=\tilde{\Psi}(s-ikv)+\tilde{\psi}(s-ikv)\hat{q}(k)\tilde{
\hat{P}}(k,s),\\
\end{equation}
we obtain the algebraic relation
\begin{equation}
\tilde{{\hat P}}(k,s)=\frac{\tilde{\Psi}(s-ikv)}{1-\tilde{\psi}(s-ikv)\hat{q}(k)}
\label{flprop_app}.
\end{equation}
Eq.~\eqref{flprop_app} is similar to the Montroll-Weiss equation \cite{montroll}  
for continuous time random walk processes. Rewriting Eq.~\eqref{flprop_app} in
terms of a geometric series, $\tilde{\hat P}(k,s)$ becomes
\begin{equation}
\tilde{{\hat P}}(k,s)=\tilde{\Psi}(s-ikv) \sum_{n=0}^{\infty}\left(\tilde{\psi}
(s-ikv) \hat{q}(k)\right)^n.
\end{equation}
With definition (\ref{Phi_n}) we end up with the compact expression
\begin{equation}
\tilde{{\hat P}}(k,s)= \sum_{n=0}^{\infty}\tilde{\Phi}_n(s-ikv) \hat{q}^n(k).
\end{equation}
Note that by definition $\Phi_n(t)$ is the probability of exactly $n$ resetting
events in $[0,t]$, and with $\int_0^t\psi_{n-1}\left(t-t'\right)\psi(t')dt'$
($\psi_0(t)=\delta(t)$, i.e., $\Phi_0(t)=\Psi(t)$), the Laplace transform of
$\Phi_n(t)$ becomes $\tilde{\Phi}_n(s)=\tilde\Psi(s)\tilde{\psi}^n(s)$. With
these relations we can perform the inverse Laplace transform of $\tilde{{\hat
P}}(k,s)$ yielding the characteristic function ${\hat P}(k,t)$
\begin{eqnarray}
\hat{P}(k,t)=\sum^\infty_{n=0}\Phi_n(t)\exp(ikvt)\hat{q}^n(k)
\label{app_char_jump}.
\end{eqnarray}

An alternative approach to derive the characteristic function is to use its 
representation as a jump diffusion process \cite{Kou},
\begin{equation}
x(t)=vt+\sum^{n(t)}_{j=1}z_j,
\end{equation}
where the stochastic variable $n(t)$ is the number of resets in the interval
$[0,t]$. The characteristic function can be computed as
\begin{eqnarray}
\nonumber
\hat{P}(k,t)&=&\langle\exp(ikx(t))\rangle\\
\nonumber
&=&\exp(ikvt)\left<\prod^{n(t)}_{j=1}\exp(ikz_j)\right>\\
&=&\sum^\infty_{n=0}\Phi_n(t)\exp(ikvt)\prod^n_{j=1}\langle\exp(ikz_j)\rangle.
\end{eqnarray}
As $n(t)$ in this expression is a stochastic variable we need to sum up the
probabilities $\Phi_n(t)$ of every possible value of $n\in\mathbb{N}$.
Furthermore, we use the property of the $z_j$ to be iid random variables,
along with the identity $\Phi_0(t)=\Psi(t)$. This leads us directly to
Eq.~\eqref{app_char_jump}.

Define now $q_n(z)$ as the distribution of the total jump size $z$ after $n$ iid
jumps with distribution $q(z)$. The relation between $q_n(z)$ and $q(z)$ is
then
\begin{equation}
q_n(z)=\left\{\begin{array}{ll}\int\limits_{-\infty}^\infty dz'q_{n-1}(z-z')
q(z')&n\ge1\\\delta(z)&n=0\end{array}\right.,
\end{equation}
and thus
\begin{equation}
\label{q_n}
{\hat q}_n(k)={\hat q}^n(k).
\end{equation}
With $q_n(z)$ from Eq.~\eqref{q_n} we take the inverse Fourier transform of the
characteristic function $\hat{P}(k,t)$, Eq.~\eqref{app_char_jump}. Thus, $P(x,t)$
takes on the form
\begin{eqnarray}
\nonumber
P(x,t)&=&\sum^\infty_{n=0}\Phi_n(t)q_n(x-vt),\\
&=&\Psi(t)\delta(x-vt)+\sum^\infty_{n=1}\Phi_n(t){q}_n(x-vt)
\label{result_indep}
\end{eqnarray}

\subsubsection*{Calculation of moments}

For the average $\langle x(t)\rangle$ and the variance $\mathrm{Var}\{x(t)\}$ of
the variable $x(t)$ we compute the first and second derivatives of $\hat{P}(k,t)$,
Eq.~\eqref{app_char_jump},
\begin{eqnarray}
\nonumber
\hat{P}'(k,t)&=&\sum^\infty_{n=0}\Phi_n(t)\exp(ikvt)\hat{q}^n(k)\left(ivt+\frac{n
\hat{q}'(k)}{\hat{q}(k)}\right) \label{indep_1dev},\\
\nonumber
\hat{P}''(k,t)&=&\sum^\infty_{n=0}\Phi_n(t)\exp(ikvt)\hat{q}^n(k)\\
&&\hspace*{-1.6cm}\times\left(\left(ivt
+\frac{n\hat{q}'(k)}{\hat{q}(k)}\right)^2+n\frac{\hat{q}''(k)\hat{q}(k)-\left(
\hat{q}'(k)\right)^2}{\left(\hat{q}(k)\right)^2}\right).
\label{indep_2dev}
\end{eqnarray}
Let $\langle z\rangle=-i\hat{q}'(0)$ be the average of the random independent
amplitude $z$ with the corresponding distribution $q(z)$. Then with
Eq.~\eqref{indep_1dev} the average $\langle x(t)\rangle$ of $x(t)$ is
\begin{equation}
\label{indep_aver}
\langle x(t)\rangle=-i\hat{P}'(0,t)=\sum^\infty_{n=0}\Phi_n(t)\big(vt+n\langle
z \rangle\big).
\end{equation}
Now let $\mathrm{Var}\{z\}=\left(\hat{q}'(0)\right)^2-\hat{q}''(0)$ be the
variance of the random independent amplitude $z$ with distribution $q(z)$.
Thus, the variance $\mathrm{Var}\{x(t)\}$ of the position $x(t)$ becomes
\begin{eqnarray}
\nonumber
\mathrm{Var}\{x(t)\}&=&\left(\hat{P}'(0,t)\right)^2-\hat{P}''(0,t),\\
&&\hspace*{-2.2cm}=\sum^\infty_{n=0}\Phi_n(t)\big((vt+n\langle z\rangle)^2+n
\mathrm{Var}\{ z\}\big)-\langle x(t)\rangle^2.
\label{indep_var}
\end{eqnarray}

\subsection{Ballistic propagation with exponential resetting amplitudes}

For the concrete choice of exponential resetting amplitudes, defined by
\begin{equation}
\label{expamp}
q(z)=\Theta(-z)\zeta^{-1}\exp\left(\frac{z}{/\zeta}\right),
\end{equation}
the distribution $q_n(z)$ becomes
\begin{eqnarray}
\nonumber
q_n(z)&=&\frac{1}{2\pi}\int_{-\infty}^\infty dk\exp(-ik z)\left(\frac{1}
{1+ik\zeta}\right)^n,\\
\nonumber
&=&\frac{(-z)^{n-1}}{\zeta^n( n-1)!}\exp\left(\frac{z}{\zeta}\right)\Theta(-z). 
\end{eqnarray}
The density $P(x,t)$ (Eq.~\eqref{result_indep}) then yielsd in the form
\begin{eqnarray}
\nonumber
P(x,t)&=&\Psi(t)\delta(x-vt)+\sum^\infty_{n=1}\Phi_n(t)\frac{(vt-x)^{n-1}}{\zeta^n
( n-1)!}\\
&&\times\exp\left(\frac{x-vt}{\zeta}\right)\Theta(vt-x).
\label{P_bal_exp}
\end{eqnarray}
The Fourier transform of $q(z)$ is $\hat{q}(k)=1/(1+ik\zeta)$. With the first
and second derivative of $\hat{q}(k)$,
\begin{equation}
\hat{q}'(k)=\frac{-i\zeta}{(1+ik\zeta)^2},\quad
\hat{q}''(k)=\frac{-2\zeta^2}{(1+ik\zeta)^3}
\end{equation}
we get the average and the variance of $z$,
\begin{eqnarray}
\nonumber
\langle z\rangle&=&-i\hat{q}'(0)=-\zeta\\
\mathrm{Var}\{z\}&=&\left(\hat{q}'(0)\right)^2-\hat{q}''(0)=-\zeta^2+2\zeta^2
=\zeta^2.
\end{eqnarray}
The mean $\langle x(t)\rangle$ (Eq.~\eqref{indep_aver}) now becomes 
\begin{equation}
\langle x(t)\rangle=\sum^\infty_{n=0}\Phi_n(t)\times(vt-n\zeta),
\label{exp_aver}
\end{equation}
and the variance $\mathrm{Var}\{ x(t)\}$ (Eq.~\eqref{indep_var}) reads
\begin{eqnarray}
\nonumber
\mathrm{Var}\{x(t)\}&=&\sum^\infty_{n=0}\Phi_n(t)\times\left( (vt-n\zeta)^2+n
\zeta^2\right)\\
&&-\left(\sum^\infty_{n=0}\Phi_n(t)\times(vt-n\zeta)\right)^2.
\label{exp_var}
\end{eqnarray}

\subsubsection*{Ballistic propagation with exponential resetting amplitude and
Poissonian resetting times}

As a specific example we consider the combination of an exponential resetting
amplitude PDF \eqref{expamp} of width $\zeta$ and Poissonian resetting times
with distribution
\begin{equation}
\label{exptime}
\psi(t)=r\exp(-rt).
\end{equation}
This implies the distributions
\begin{equation}
\tilde{\psi}(s)=\frac{r}{r+s},\,\,\,\tilde{\Psi}(s)=\frac{1-{\tilde{\psi}}(s)}{s}
=\frac{1}{r+s},
\end{equation}
and from this expression we find the Laplace transform
\begin{equation}
\tilde{\Phi}_n(s)=\tilde{\Psi}(s)\tilde{\psi}^n(s)=\frac{1}{r}\left( 
\frac{r}{r+s} \right)^{n+1}.
\end{equation}
After Laplace inversion,
\begin{equation}
\label{Pois_Phi}
\Phi_n(t)=\frac{(rt)^n}{n!}\exp(-rt).
\end{equation}
This yields the density $P(x,t)$ (Eq.~\eqref{P_bal_exp}) for this case,
\begin{eqnarray}
\nonumber
P(x,t)&=&\exp(-rt)\delta(x-vt)+\sum^\infty_{n=1}\frac{(rt)^n(vt-x)^{n-1}}
{\zeta^nn!(n-1)!}\\
&&\times\exp\left(\frac{x-vt-rt\zeta}{\zeta}\right)\Theta(vt-x).
\end{eqnarray}
With the representation
\begin{equation}
I_1(\xi)=\frac{\xi}{2}\sum^\infty_{n=0}\frac{\left(\xi^2/4\right)^n}{n!(n+1)!}
\end{equation}
of the modified Bessel function of the first kind we then obtain our result,
\begin{eqnarray}
\nonumber
P(x,t)&=&e^{-rt}\delta(x-vt)+\exp([x-vt-rt\zeta]/\zeta)\\
&&\hspace*{-1.2cm}
\times\sqrt{\frac{rt/\zeta}{vt-x}}I_1\left(2\sqrt{\frac{rt}{\zeta}(vt-x)}\right)
\Theta(vt-x).
\label{P_bal_exp_Pois}
\end{eqnarray}

The mean $\langle x(t)\rangle$ of $x(t)$ (Eq.~\eqref{exp_aver}) is
\begin{eqnarray}
\nonumber
\langle x(t)\rangle&=&\sum^\infty_{n=0}\frac{(rt)^n}{n!}\exp(-rt)\times
(vt-n\zeta),\\
&=&(v-r\zeta)t.
\label{exp_Pois_aver}
\end{eqnarray}
The mean position thus depends linearly on $t$ and increases or decreases,
depending on the sign of $(v-r\zeta)$. The variance $\mathrm{Var}\{x(t)\}$
(Eq.~\eqref{exp_var}) has the form
\begin{eqnarray}
\nonumber
\mathrm{Var}\{x(t)\}&=&\sum^\infty_{n=0}\frac{(rt)^n}{n!}\exp(-rt)\times\left( 
(vt-n\zeta)^2+n \zeta^2\right)\\
\nonumber
&&-(vt-rt\zeta)^2,\\
&=&2rt\zeta^2.
\label{exp_Pois_var}
\end{eqnarray}
The variance is thus also proportional to $t$, but it is $v$-independent.

\begin{figure}
\includegraphics[width=8.8cm]{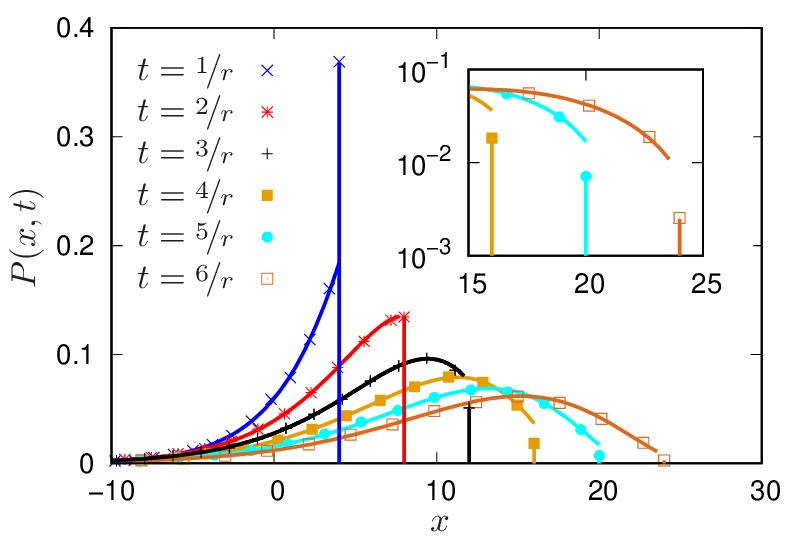}
\caption{Height profile PDF $P(x,t)$ as function of $x$ for six different $t$
for ballistic motion with Poissonian resetting times and exponential resetting
amplitudes. The probability of no reset until $t$ is represented by the vertical
line at $x=vt$, it is shown on log-lin scale for
different $t$ in the inset. Simulations results are shown by points, the
analytical results are shown by solid lines. Parameters: $v=0.5$, $r=0.125$,
and $\zeta=2$.}
\label{pdf_indep_poissonian_reset}
\end{figure}

Fig.~\ref{pdf_indep_poissonian_reset} shows $P(x,t)$ at different times: the
maximum value decreases and the PDF gradually shifts away from negative values.
The possibility of no reset up to time $t$ is encoded in the finite value at
$x=vt$, the inset shows a discontinuity of $P(x,t)$ at $x=vt$ and the exponential
relation between the probability of no reset and time $t$.

In App.~\ref{different} we derive the Fourier transform of the PDF $P(x,t)$
from the master equation formulation for the case of ballistic propagation,
Poissonian resetting times and arbitrary independent resetting amplitudes.
The result (\ref{another_approach_indep_char_sol}) then corresponds to
Eq.~(\ref{app_char_jump}) with the choice (\ref{Pois_Phi}) for $\Phi_n(t)$.

\subsection{Ballistic displacement with constant pace and exponential 
resetting amplitudes}

We now consider another variant of ballistic propagation, namely, of a constant
duration between successive resetting events, which we refer to as constant
pace. The distribution of the resetting interval lengths is
\begin{equation}
\psi(t)=\delta\left(t-\frac{1}{r}\right).
\end{equation}
In Laplace space this implies the distributions
\begin{equation}
\label{copadis}
\tilde{\psi}(s)=\exp\left(-\frac{s}{r}\right),\,\,\,\tilde{\Psi}(s)=\frac{1-\exp
(-s/r)}{s},\\
\end{equation}
and consequently
\begin{equation}
\tilde{\Phi}_n(s)=\frac{\exp(-ns/r)-\exp(-(n+1)s/r)}{s},
\end{equation}
After Laplace inversion,
\begin{equation}
\Phi_n(t)=\Theta\left(t-\frac{n}{r}\right)-\Theta\left(t-\frac{n+1}{r}\right).
\label{const_Phi}
\end{equation}
Thus, the density $P(x,t)$ (Eq.~\eqref{P_bal_exp}) is given by
\begin{eqnarray}
\nonumber
P(x,t)&=&\left(\Theta(t)-\Theta\left( t-\frac{1}{r}\right)\right)\delta(x-vt)\\
\nonumber
&&+\Theta(vt-x)\sum^\infty_{n=1}\frac{(vt-x)^{n-1}}{\zeta^n(n-1)!}\exp\left(
\frac{x-vt}{\zeta}\right)\\
&&\times\left(\Theta\left(t-\frac{n}{r}\right)-\Theta\left(t-\frac{n+1}{r}
\right)\right).
\label{density}
\end{eqnarray}

The mean $\langle x(t)\rangle$ of $x(t)$ (Eq.~\eqref{exp_aver}) becomes 
\begin{eqnarray}
\nonumber
\langle x(t)\rangle&=&\sum^\infty_{n=0}\left(\Theta\left(t-\frac{n}{r}\right)
-\Theta\left(t-\frac{n+1}{r}\right)\right)\\
\nonumber
&&\times(vt-n\zeta)\\
&=&vt-\zeta\lfloor rt\rfloor
\label{exp_const_aver},
\end{eqnarray}
where we introduce the floor function $\lfloor x\rfloor=\max\{l\in\mathbb{Z}|l
\le x\}$. The variance $\mathrm{Var}\{ x(t)\}$ (Eq.~\eqref{exp_var}) reads
\begin{eqnarray}
\nonumber
\mathrm{Var}\{x(t)\} &=&\zeta^2\sum^\infty_{n=0}\left(\Theta\left(t-\frac{n}{r}
\right)-\Theta\left(t-\frac{n+1}{r}\right)\right)n,\\
&=&\zeta^2\lfloor rt\rfloor.
\label{exp_const_var}
\end{eqnarray}
In the long time limit results \eqref{exp_const_aver} and \eqref{exp_const_var}
coincide with the corresponding mean and variance in the Poissonian resetting
time scenario, Eqs.~\eqref{exp_Pois_aver} and \eqref{exp_Pois_var}.

\begin{figure}
\includegraphics[width=8.8cm]{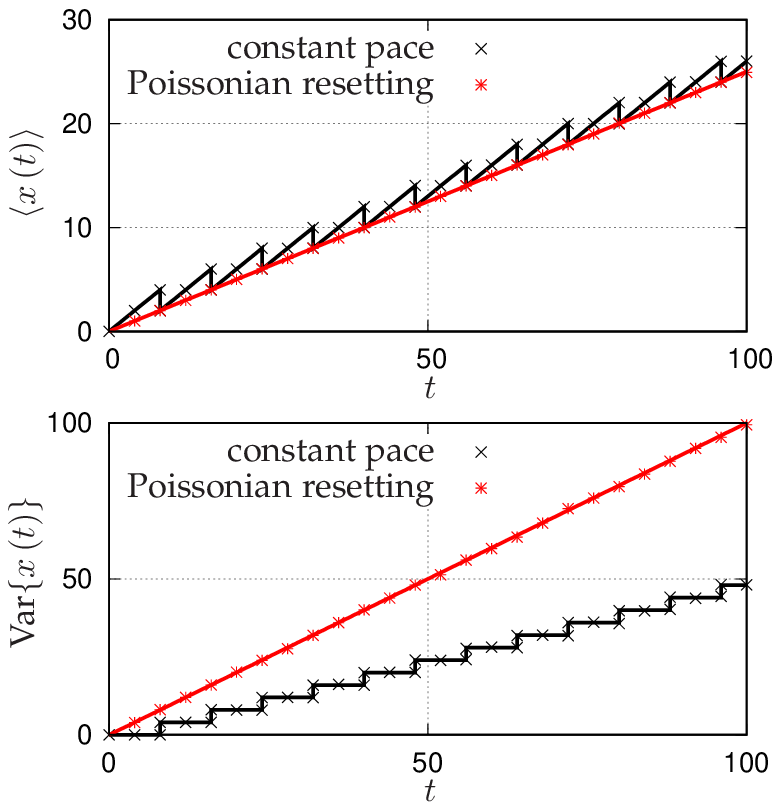}
\caption{Average and variance of the height $x$ as function for $t$ for
exponential resetting amplitudes and ballistic motion, with Poissonian
and constant pace resetting times. Points represent simulations results,
solid lines are the analytical results. Parameters: $v=0.5$, $r=0.125$, and
$\zeta=2$.}
\label{moments_ind}
\end{figure}

In Fig.~\ref{moments_ind} the mean position and variance are shown for two
different examples of ballistic propagation and exponential resetting amplitudes,
demonstrating the linear growth of the mean height. In this example we see that
the constant pace scenario has the same mean as the Poissonian resetting model
but half the variance, as can also be seen from comparison of
Eqs.~(\ref{exp_Pois_var}) and (\ref{exp_const_var}).

\begin{figure*}
\includegraphics{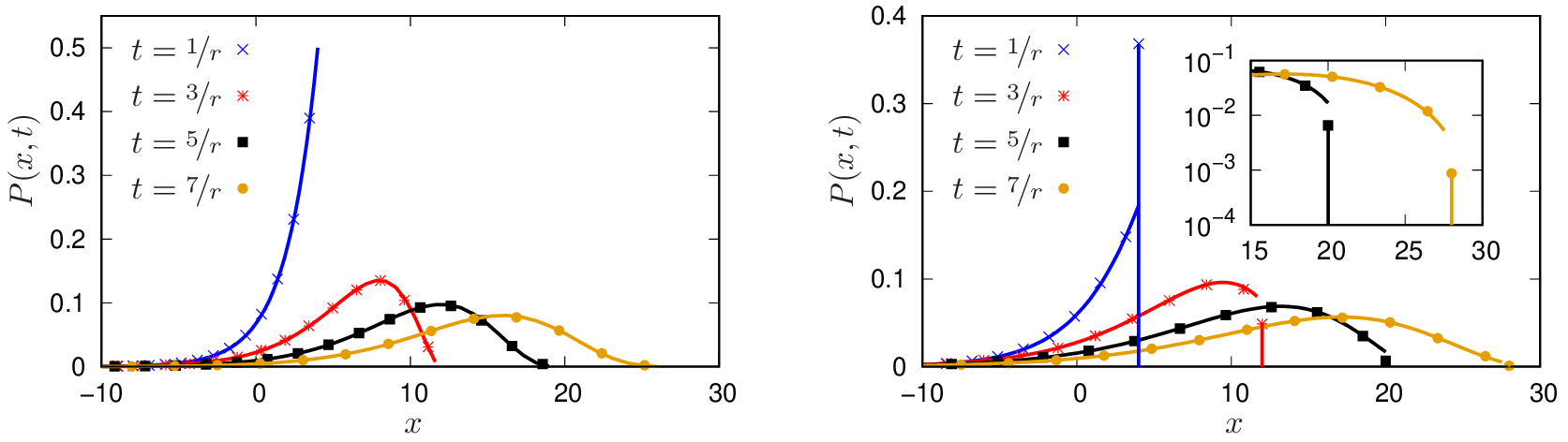}
\caption{Height profile $P(x,t)$ as function of $x$ for four different $t$, for
ballistic motion with exponential resetting amplitudes and two different 
resetting scenarios. Left: "constant pace" resetting. Right: Poissonian 
resetting. The probability of no reset until $t$ is represented by the vertical 
line at $x=vt$, it is shown on log-lin scale for different $t$ in the inset of 
the right panel. Points: simulations, Solid lines: analytical results.
Parameters: $v=0.5$, $r=0.125$, and $\zeta=2$.}
\label{fig1_app}
\end{figure*}

Let us compare the difference between the cases of constant pace and Poissonian
resetting intervals in more detail.
Fig.~\ref{fig1_app} illustrates the PDF $P(x,t)$ for constant pace (left panel)
and Poissonian resetting (right panel) at different times. For the chosen values
the maximum of the PDF increases with time, and the standard deviation of the
PDF increases in both panels. In the case of constant pace resetting, we showed
the distribution immediately after resetting in Fig.~\ref{fig1_app}. For
Poissonian resetting the possibility that no reset occurs up to time $t$ is
encoded in the finite value at $x = vt$. Its value is detailed in the inset,
showing a discontinuity of $P(x,t)$ at $x=vt$ and the exponential relation
between the probability of no reset and time $t$.

\begin{figure*}
\includegraphics{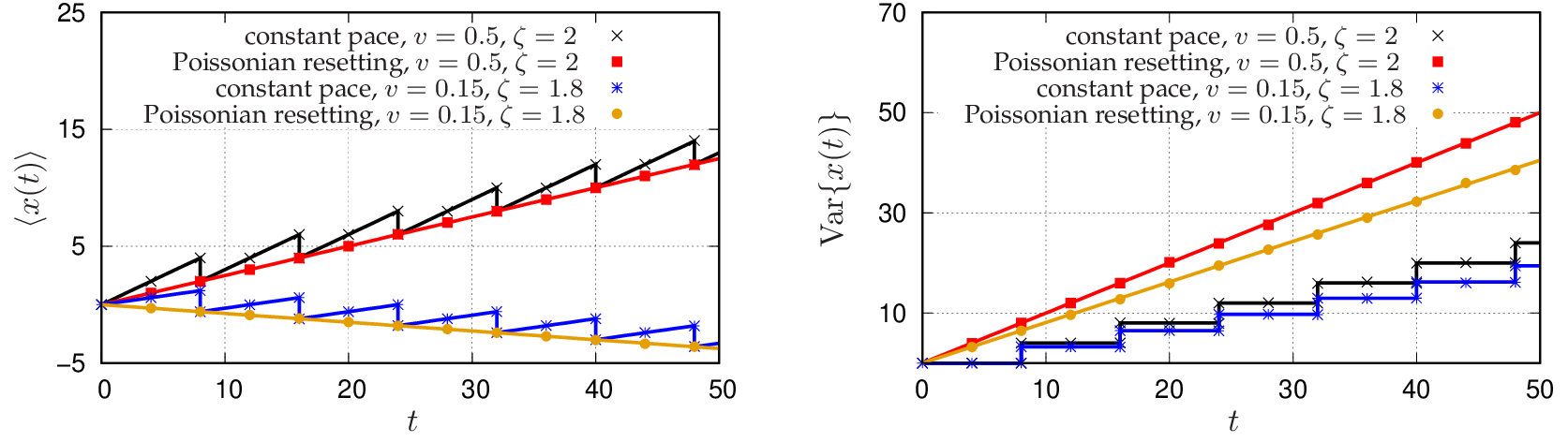}
\caption{Average and variance of the height $x$ as function of time $t$ for
exponential resetting amplitudes and ballistic motion, with Poissonian and
"constant pace" resetting. Points represent simulations, solid lines
the analytical results. For all realizations,
$r=0.125$.}
\label{fig2_app}
\end{figure*}

Fig.~\ref{fig2_app} shows the behavior of the average (left panel) and variance 
(right panel) of $x(t)$. For constant pace resetting the average $\langle x(t) 
\rangle$ increases linearly in time between successive resetting events, however 
the variance of $x(t)$ does not change in this time span. The corresponding PDF 
moves linearly in time, but does not change its shape during these time spans.  
The shape of the distribution only change at the resetting events. As it can be 
seen in Fig.~\ref{fig2_app} the variance $\mathrm{Var}\{x (t) \}$ only increases
at these times. For Poissonian resetting the mean position depends linearly on 
$t$ and increases or decreases, depending on the sign of $(v-r \zeta)$. Both 
possibilities are shown in Fig.~\ref{fig2_app}. Moreover, in presence of 
constant pace resetting, we can see that $\langle x(t) \rangle$ increases
faster than for Poissonian resetting during the resetting interval lengths. However, 
under the same choice of parameter the mean for constant pace resetting 
coincides with the Poissonian resetting at the resetting events. For 
Poissonian resetting the relation between $\mathrm{Var}\{x (t) \}$ and $t$ is 
linear and increases faster as for "constant pace" resetting.

\section{Dependent resetting picture}
\label{dep}

In many realistic situations the height $x(t)$ cannot assume negative values,
e.g., when the deposits in a river bed shrink until they reach a solid bedrock,
when the value of a given stock becomes zero, or when a population goes extinct.
Random-amplitude resetting processes with strictly positive $x$ in our framework
are described by \emph{dependent resetting amplitudes}, the main novel feature
introduced in this work.

For such dependent resetting amplitudes we use the following relation between
consecutive resetting points,
\begin{eqnarray}
\label{dep_iter_app}
\left\{\begin{array}{ll}x(t_{n+1})=(y(t_{n+1})|x(t_n))\times c_{n+1}\\
x(t_0)=x_0\end{array}\right.,
\end{eqnarray}
where the $c_n\in[0,1)$ are iid random variables of the running index $n$.
For $n=0$, Eq.~\eqref{dep_iter_app} yields
\begin{equation}
\label{dep_iter_1_app}
x(t_1)=(y(t_1)|x_0)\times c_1.
\end{equation}
With Eq.~\eqref{dep_iter_1_app}, $\phi_1(x_1,t_1;x_0,t_0)$ becomes
\begin{equation}
\label{dist_x_1_dep}
\phi_1(x_1,t_1;x_0,t_0)=\int_0^\infty\frac{dy}{y}G(y,t_1;x_0,t_0)f_C\left(\frac{
x_1}{y}\right).
\end{equation} 
In Eq.~\eqref{dist_x_1_dep} we only allow movement for positive heights, $0\le
y<\infty$. Due to our requirement that height $x(t)$ cannot assume negative 
values, we impose the additional condition that $f_C(c_n)=0$ for $c_n<0$ and
$c_n\ge1$, such that we only have to consider the range $0\le c_1=x_1/y<1$, 
in which $f_C(c_1)\neq0$. Thus we have the inequality $0\le x_1/y<1$, or
\begin{equation}
\label{cond_dep_0_1}
0\le x_1<y.
\end{equation} 
For dependent resetting amplitudes we get the \emph{first resetting picture\/} of
the process if we substitute $\phi_1(x_1,t_1;x_0,t_0)$, Eq.~\eqref{dist_x_1_dep},
into Eq.~\eqref{p_r_general_app} and considering the range of $x_1$ for which
$f_C(x_1/y)\neq0$ (compare Eq.~\eqref{cond_dep_0_1}). Thus, we get
\begin{eqnarray}
\nonumber
P(x,t;x_0,t_0)&=&\Psi(t-t_0)G(x,t;x_0,t_0)\\
\nonumber
&&\hspace{-1.8cm}+\int_{t_0}^tdt_1\psi(t_1-t_0)\int_0^{\infty}\frac{dy}{y}G(y,
t_1;x_0,t_0)\\
&&\hspace*{-1.2cm}
\times\int_0^ydx_1f_C(x_1/y)P(x,t;x_1,t_1).
\label{first_dep}
\end{eqnarray}
The key difference to Eq.~(\ref{first}) is that the $y$-integration is restricted
to $y\in[0,\infty)$ and that the resetting length PDF $q(x_1-y)$ is replaced by
the scaling function $y^{-1}f_C(x_1/y)$, that in turn is part of the product
distribution (\ref{dist_x_1_dep}). We derive the last resetting picture
corresponding to the first resetting picture (\ref{first_dep}) in
App.~\ref{dep_equal}. We note that when the PDF $G$ is homogenous in space and
time, the PDF $P$ is still homogeneous in time but the spatial homogeneity is
lost (App.~\ref{dep_equal}).

\subsection{Reduction to classical stochastic resetting}

Before proceeding with our analysis we stop to proof that our RASR process with
dependent resetting amplitudes is a generalization of classical SR. In fact we
can prove this equivalence for both the first resetting picture and the last
resetting picture if we set $f_C(c_n)=\delta(c_n)$ and use Poissonian resetting
$\psi(t)=r\exp(-rt)$ along with the initial position $x_0=0$. With this
deterministic resetting mechanism we can verify the results of \cite{RIL_3}
for the first renewal picture and \cite{Evans2} for the last renewal picture of
SR.

In the first resetting picture we have in our framework
\begin{eqnarray}
\nonumber
P(x,t;0,0)&=&\exp(-rt)G(x,t;0,0)+\int\limits_0^tdt_1r\exp(-rt_1)\\
\nonumber
&&\hspace*{-1.2cm}\times\int\limits_0^\infty \frac{dy}{y}G(y,t_1;0,0)\int
\limits_0^ydx_1\delta\left(\frac{x_1}{y}\right)P(x,t;x_1,t_1)\\
\nonumber
&&\hspace*{-1.6cm}=\exp(-rt)G(x,t;0,0)+r\int\limits_0^tdt_1\exp(-rt_1)\\
\nonumber
&&\hspace*{-1.2cm}\int\limits_0^\infty dyG(y,t_1 ;0,0)\int\limits_0^1dc_1
\delta(c_1)P(x,t;c_1y,t_1),
\end{eqnarray}
in which $c_1=x_1/y$. This implies that
\begin{eqnarray}
\nonumber
P(x,t;0,0)&=&\exp(-rt)G(x,t;0,0)+r\int\limits_0^tdt_1\exp(-rt_1)\\
\nonumber
&&\times\int\limits_0^\infty dyG(y,t_1;0,0)P(x,t;0,t_1)\\
\nonumber
&=&\exp(-rt)G(x,t;0,0)+r\int\limits_0^tdt_1\exp(-rt_1)\\
&&\times P(x,t;0,t_1),
\end{eqnarray}
and therefore proves the equivalence to \cite{RIL_3} with $x_0=0$.

\begin{widetext}
Conversely, in the last resetting picture we have [cf.~Eq.~(\ref{dep_in_homogen_app})]
\begin{eqnarray}
\nonumber
P(x,t;0,0)&=&\exp(-rt)G(x,t;0,0)+\sum^\infty_{n=1}\int\limits_0^td\tau_n\int
\limits_0^1dc_n\int\limits_0^\infty dy'_n\left(\prod^{n-1}_{i=1}\int\limits
_0^{\tau_{n+1-i}}d\tau_{n-i}r\exp(-r(\tau_{n+1-i}-\tau_{n-i}))\right)\\
\nonumber
&&\times\left(\prod^{n-1}_{i=1}\int\limits_0^\infty dy'_{n-i} G(y'_{n+1-i},
\tau_{n+1-i};c_{n-i}y'_{n-i},\tau_{n-i})\int\limits_0^1dc_{n-i}\delta(c_{n+1-i})
\right)\\
\nonumber
&&\times\delta(c_1)r\exp(-r\tau_1)G(y'_1,\tau_1;0,0)\exp(-r(t-\tau_n))G(x,t;
c_ny'_n,\tau_n)\\
\nonumber
&=&\exp(-rt)G(x,t;0,0)+\sum^\infty_{n=1}r^n\int\limits_0^td\tau_n\left(\prod
^{n-1}_{i=1}\int\limits_0^{\tau_{n+1-i}}d\tau_{n-i}\right)\exp(-r[\tau_n-
\tau_{n-1}])\exp(-r[\tau_{n-1}-\tau_{n-2}])\\
\nonumber
&&\times\ldots\times\exp(-r[\tau_3-\tau_2])\exp(-r[\tau_2-\tau_1])\exp(-r\tau_1)
\exp(-r[t-\tau_n])G(x,t;0,\tau_n)\\
&=&\exp(-rt)G(x,t;0,0)+r\int\limits_0^t d\tau\sum^\infty_{n=1}\frac{(r\tau)^{
n-1}}{(n-1)!}\exp(-rt)G(x,t;0,\tau),
\end{eqnarray}
with $\tau=\tau_n$. This demonstrates that
\begin{equation}
P(x,t;0,0)=\exp(-rt)G(x,t;0,0)+r\int\limits_0^td\tau\exp(-r[t-\tau])G(x,t;0,\tau),
\end{equation}
and completes our proof of equivalence with the formulation in \cite{Evans2} for 
$x_0=0$.

\subsection{Ballistic propagation with dependent resetting amplitude}

For the spatial Laplace transform $\bar{P}(u,t;x_0)=\int_0^\infty dx\exp(-ux)
P(x,t;x_0)$ of the one-sided density $P(x,t;x_0)=P(x,t;x_0,t_0=0)$ in the first
resetting picture (\ref{first_dep}) respectively the last resetting picture
\eqref{p_r_dep_last_app} for the case of ballistic propagation, we use
Eq.~\eqref{p_r_dep_last_app} with $G(x,t;y,\tau)=\delta(x-y-v(t-\tau))$.
Collecting terms, $P(x,t;x_0)$ reads
\begin{eqnarray}
\nonumber
P(x,t;x_0)&=&\Psi(t)\delta(x-x_0-vt)+\sum^\infty_{n=1}\int\limits_0^td\tau_n
\int\limits_0^1dc_n\int\limits_0^\infty dy_n
\times\left(\prod^{n-1}_{i=1}\int\limits_0^{\tau_{n+1-i}} d\tau_{n-i}\psi(\tau
_{n+1-i}-\tau_{n-i})\right.\\
\nonumber
&&\left.\times\int\limits_0^\infty dy_{n-i}\delta(y_{n+1-i}-c_{n-i}y_{n-i}-v
(\tau_{n+1-i}-\tau_{n-i}))\int\limits_0^1 dc_{n-i}f_C(c_{n+1-i})\right)\\
&&\times f_C(c_1)\psi(\tau_1)\delta(y_1-x_0-v\tau_1)\Psi(t-\tau_n)
\delta(x-c_ny_n-v(t-\tau_n)),
\end{eqnarray}
and after the spatial Laplace transform we find
\begin{eqnarray}
\nonumber
\bar{P}(u,t;x_0)&=&\Psi(t)\exp(-u(x_0+vt))+\sum^\infty_{n=1}\int\limits_0^td
\tau_n\int\limits_0^1dc_n\left(\prod^{n-1}_{i=1}\int\limits_0^{\tau_{n+1-i}}
d\tau_{n-i}\psi(\tau_{n+1-i}-\tau_{n-i})\int\limits_0^1 dc_{n-i}f_C(c_{n+1-i}
)\right)\\
&&\times f_C(c_1)\psi(\tau_1)\Psi(t-\tau_n)\exp\left(-u\left(x_0\prod^n_{j=0}c_j+ 
v(t-\tau_n)+v \sum^n_{j=1}(\tau_j-\tau_{j-1})\prod^n_{k=j} c_k \right)\right),
\label{app_dep_bal_lap}
\end{eqnarray}
in which $c_0=1$ and $\tau_0=0$. Performing a Laplace transform in time (with
the corresponding Laplace variable $s$), in addition, our general result for
the PDF reads
\begin{eqnarray}
\tilde{\bar{P}}(u,s;x_0)&=&\sum^\infty_{n=0}\tilde\Psi(s+uv)\left(\prod^n_{k=1}
\int\limits_0^1dc_kf_C(c_k)\tilde\psi\left(s+uv\prod^k_{i=1}c_i\right)\right)
\exp\left(-ux_0\prod^n_{j=0}c_j\right).
\label{app_dep_bal_lap_lap}
\end{eqnarray}

To compute the mean
\begin{equation}
\langle x(t)|x_0\rangle=-{\bar{P}}'(0,t;x_0)
\label{app_dep_bal_aver}
\end{equation}
and variance
\begin{equation}
\mathrm{Var}\{ x(t)|x_0\}={\bar{P}}''(0,t;x_0)-\left({\bar{P}}'(0,t;x_0)\right),
\label{app_dep_bal_var}
\end{equation}
we use the first and second derivatives of $\bar{P}(u,t;x_0)$,
Eq.~(\ref{app_dep_bal_lap}), with respect to $u$ and set $u=0$.
It is easier to work with the Laplace transform (\ref{app_dep_bal_lap_lap}) in time.
General formulas for the first and second derivatives of
Eq.~(\ref{app_dep_bal_lap_lap}) with respect to the Laplace variable $u$ are
presented in App.~\ref{appe}. They will be used in Sections \ref{sec4c} and
\ref{sec4d} below.

\subsection{Ballistic displacement with arbitrary resetting times and
uniform dependent resetting amplitudes}
\label{sec4c}

We now turn to the ballistic displacement process with arbitrary resetting
intervals but the specific choice of uniform dependent resetting amplitudes.
This choice allows us to specify \eqref{app_dep_bal_lap_lap_first_der_zero}
and \eqref{app_dep_bal_lap_lap_second_der_zero} when we inlude $f_C(c)=1$.
Thus for the first and second moment of $c$ we get $\langle c\rangle=1/2$ and
$\langle c^2\rangle=1/3$. The first derivative $\tilde{\bar{P}}'(u,t;x_0)$ becomes 
\begin{equation}
\tilde{\bar{P}}'(0,s;x_0)=\sum^\infty_{n=0}\left(v\left(\tilde{\psi}^n(s)\tilde{
\Psi}'(s)+\tilde{\psi}^{n-1}(s)\tilde{\psi}'(s)\tilde{\Psi}(s)\left(1-\frac{1}{2^n}
\right)\right)-x_0\left(\frac{\tilde{\psi}(s)}{2}\right)^n\tilde{\Psi}(s)\right) 
\label{app_uni_first_der}.
\end{equation}
The second derivative $\tilde{\bar P}''(u,t;x_0)$ reads
\begin{eqnarray}
\nonumber
\tilde{\bar{P}}''(0,s;x_0)&=&\sum^\infty_{n=0}v^2\left(\tilde{\psi}^n(s)\tilde{
\Psi}''(s)+\frac{1}{2}\tilde{\psi}^{n-1}(s)\tilde{\psi}''(s)\tilde{\Psi}(s)\left(
1-\frac{1}{3^n}\right)+2\tilde{\psi}^{n-1}(s)\tilde{\psi}'(s)\tilde{\Psi}'(s)
\left(1-\frac{1}{2^n}\right)\right)\\
\nonumber
&&+\sum^\infty_{n=0}\left(v^2\tilde{\Psi}(s)\tilde{\psi}'^2(s)\tilde{\psi}^{n-2}(s)
\left(1+\frac{3}{3^n}-\frac{4}{2^n}\right)+x^2_0\left(\frac{\tilde{\psi}(s)}{3}
\right)^n\tilde{\Psi}(s)\right)\\
&&-\sum^\infty_{n=0}2vx_0\left(\left(\frac{\tilde{\psi}(s)}{2}\right)^n\tilde{
\Psi}'(s)+2\tilde{\psi}^{n-1}(s)\tilde{\psi}'(s)\tilde{\Psi}(s)\left(\frac{1}{
2^n}-\frac{1}{3^n}\right)\right).
\label{app_uni_second_der}
\end{eqnarray}

For \emph{constant pace resetting times},
we have a periodic reset with $\psi(t)=\delta\left(t-1/r\right)$
corresponding to expressions \eqref{copadis}. Thus, the resetting amplitude is
the only stochastic variable in this process. After some algebra and Laplce
inversion we find
\begin{eqnarray}
\bar{P}'(0,t;x_0)=-\sum_{n=0}^\infty\Phi_n(t)\left(v\left(t-\frac{n}{r}\right)
+\frac{v}{r}\left(1-\frac{1}{2^n}\right)+\frac{x_0}{2^n}\right),
\end{eqnarray}
in which $\Phi_n(t)=\Theta\left(t-n/r \right)-\Theta\left(t-(n+1)/r \right)$.
The mean $\langle x(t)|x_0\rangle$, Eq.~\eqref{app_dep_bal_aver}, then yields
in the form
\begin{eqnarray}
\langle x(t)|x_0\rangle =x_0+vt+\sum^{\lfloor rt\rfloor}_{n=1}\left(
\frac{1}{2^n}\left( \frac{v}{r}-x_0\right)-\frac{v}{r}\right)
\label{aver_bal_uni_const}
\end{eqnarray}
with the asymptotic properties
\begin{eqnarray}
\nonumber
&&\limsup_{t\to\infty}\langle x(t)|x_0\rangle=2\frac{v}{r},\\
&&\liminf_{t\to\infty} \langle x(t)|x_0\rangle=\frac{v}{r}.
\label{newbounds}
\end{eqnarray}
Thus, in the long time limit the oscillating average $\langle x(t)|x_0\rangle$
is restricted by the two bounds (\ref{newbounds}).

Similarly we compute the second derivative of the PDF,
\begin{eqnarray}
\nonumber
\bar{P}''(0,t;x_0)&=&\sum_{n=0}^\infty \Phi_n(t)\left(v^2\left(t-
\frac{n}{r}\right)^2+\frac{2v^2}{r}\left(t-\frac{n}{r}\right)\left(1-\frac{1}{2^n}
\right)+\frac{v^2}{2r^2}\left( 3+\frac{5}{3^n}-\frac{8}{2^n}\right)-\frac{2vx_0}{2^n}
\left(t-\frac{n}{r}\right)\right)\\
&&+\sum_{n=0}^{\infty}\Phi_n(t)\left(\frac{4vx_0}{r}\left(\frac{1}{2^n}-\frac{1}{3^n}
\right)+\frac{x_0^2}{3^n} \right),
\end{eqnarray}
in which $\Phi_n(t)=\Theta\left(t-n/r \right)-\Theta\left(t-(n+1)/r \right)$.
The variance, Eq.~\eqref{app_dep_bal_var}, finally reads
\begin{eqnarray}
\nonumber
\mathrm{Var}\{x(t)|x_0\}&=&\sum^{\lfloor rt\rfloor}_{n=1}\left(x^2_0
\left(\frac{3}{4^n}-\frac{2}{3^n}\right)+2\frac{x_0 v}{r}\left(\frac{4}{3^n}-\frac{
3}{4^n}-\frac{1}{2^n}\right)+\frac{1}{2}\left(\frac{v}{r}\right)^2\left(\frac{6}{4
^n}+\frac{4}{2^n}-\frac{10}{3^n}\right)\right)\\
&\stackrel{t\to\infty}{\longrightarrow}&\frac{1}{2}\left(\frac{v}{r}\right)^2.
\label{var_bal_uni_const}
\end{eqnarray}

\subsection{Ballistic propagation and Poissonian resetting times}
\label{sec4d}

We now consider Poissonian resetting intervals with rate $r$, $\psi(t)=r\exp(-rt)$.
Such exponential distributions are in fact used in several SR studies, including
\cite{Evans1,RM_del3,RIL_5,SC_1}. For the resetting amplitudes we first derive
a general solution and then consider specific examples.

We start from Eqs.~\eqref{app_dep_bal_lap_lap_first_der_zero} and
\eqref{app_dep_bal_lap_lap_second_der_zero} and use the resetting
time distributions with their Laplace transforms $\tilde{\psi}(s)=r/(r+s)$ and
$\tilde{\Psi}(s)=1/(r+s)$. Evaluating the geometric series we obtain the
derivatives of the PDF. After Laplace inversion, these read
\begin{eqnarray}
\label{app_mom_dep_first}
\bar{P}'(0,t;x_0)&=&\frac{v}{r\left(1-\langle c\rangle\right)}\left(\exp\left(
-rt\left(1-\langle c\rangle\right)\right)-1\right)-x_0\exp\left(-rt\left(1
-\langle c\rangle\right)\right),\\
\nonumber
\bar{P}''(0,t;x_0)&=&\frac{2v^2}{r^2\left(\langle c\rangle-\langle c^2\rangle
\right)}\left(\frac{\exp\left(-rt\left(1-\langle c^2\rangle\right)\right)}{1-
\langle c^2\rangle}-\frac{\exp\left(-rt\left(1-\langle c\rangle\right)\right)}{
1-\langle c\rangle}\right)\\
\nonumber
&&+\frac{2v^2}{r^2\left(1-\langle c\rangle\right)\left(1-\langle c^2\rangle\right)}
+x^2_0\exp(-rt[1-\langle c^2\rangle])\\
&&+\frac{2x_0v}{r\left(\langle c\rangle-\langle c^2\rangle\right)}\left(\exp(-rt
[1-\langle c\rangle])-\exp(-rt[1-\langle c^2\rangle])\right).
\label{app_mom_dep_second}
\end{eqnarray}
We then derive the mean and variance,
\begin{eqnarray}
\label{app_dep_average}
\langle x(t)|x_0\rangle&=&\frac{v}{r\left(1-\langle c\rangle\right)}\left(1-
\exp(-rt[1-\langle c\rangle])\right)+x_0\exp(-rt[1-\langle c\rangle]),\\
\nonumber
\mathrm{Var}\{x(t)|x_0\}&=&\frac{2v^2\exp(-rt)}{r^2(\langle c\rangle-\langle
c^2 \rangle)}\left(\frac{\exp(rt\langle c^2\rangle)}{1-\langle c^2\rangle}-
\frac{\exp(rt\langle c\rangle)}{1-\langle c\rangle}\right)\\
\nonumber
&&+\frac{2v^2}{r^2\left(1-\langle c\rangle\right)}\left(\frac{1}{1-\langle c^2
\rangle}+\frac{\exp(-rt[1-\langle c\rangle])}{1-\langle c\rangle}\right)-
\frac{v^2\left(1+\exp(-2rt[1-\langle c\rangle])\right)}{r^2\left(1-\langle c
\rangle\right)^2}\\
\nonumber
&&+\frac{2x_0v\exp(-rt)}{r}\times\left(\frac{\exp(rt\langle c \rangle)-\exp(rt
\langle c^2\rangle)}{\langle c\rangle-\langle c^2 \rangle}+\frac{\exp(-rt[
1-2\langle c\rangle])-\exp(rt\langle c\rangle)}{1-\langle c\rangle}\right)\\
&&+x^2_0\left(\exp(-rt[1-\langle c^2\rangle])-\exp(-2rt[1-\langle c\rangle])
\right),
\label{app_dep_variance}
\end{eqnarray}
with the initial condition $x(0)=x_0$.

For \emph{uniformly distributed resetting amplitudes\/} with $\langle c\rangle
=1/2$ and $\langle c^2\rangle=1/3$ we then find the specific expressions
\begin{eqnarray}
\langle x(t)|x_0\rangle=x_0\exp\left(-\frac{rt}{2}\right)+2\frac{v}{r}
\left(1-\exp\left(-\frac{rt}{2}\right)\right)\stackrel{t\to\infty}{\longrightarrow}
2\frac{v}{r}
\end{eqnarray} 
and the variance
\begin{eqnarray}
\nonumber
\mathrm{Var}\{ x(t)|x_0\}&=&x^2_0\left(\exp\left(-\frac{2rt}{3}\right) -\exp(-rt)
\right)+\frac{vx_0}{r}\left( 4\exp(-rt)+8\exp\left(-\frac{rt}{2}\right)-12\exp
\left(-\frac{2rt}{3} \right)\right)\\
&&+\left(\frac{v}{r}\right)^2\left(2-16\exp\left(-\frac{rt}{2}\right)+18\exp\left(
-\frac{2rt}{3}\right)-4\exp(-rt)\right)\stackrel{t\to\infty}{\longrightarrow}
2\left(\frac{v}{r}\right)^2.
\label{var_bal_uni_exp}
\end{eqnarray}

Moreover, for the case of a \emph{deterministic reset to the initial height},
$\langle c\rangle=0$ and $\langle c^2\rangle=0$, we arrive at
\begin{eqnarray}
\label{app_classical_reset_average}
&&\langle x(t)|x_0=0\rangle=\frac{v}{r}(1-\exp(-rt))\\
&&\mathrm{Var}\{x(t)|x_0=0\}=\frac{v^2}{r^2}-\frac{2v^2t\exp(-rt)}{r}
-\frac{v^2\exp(-2rt)}{r^2}.
\label{app_classical_reset_variance}
\end{eqnarray}

\begin{figure}
\includegraphics[width=6.8cm]{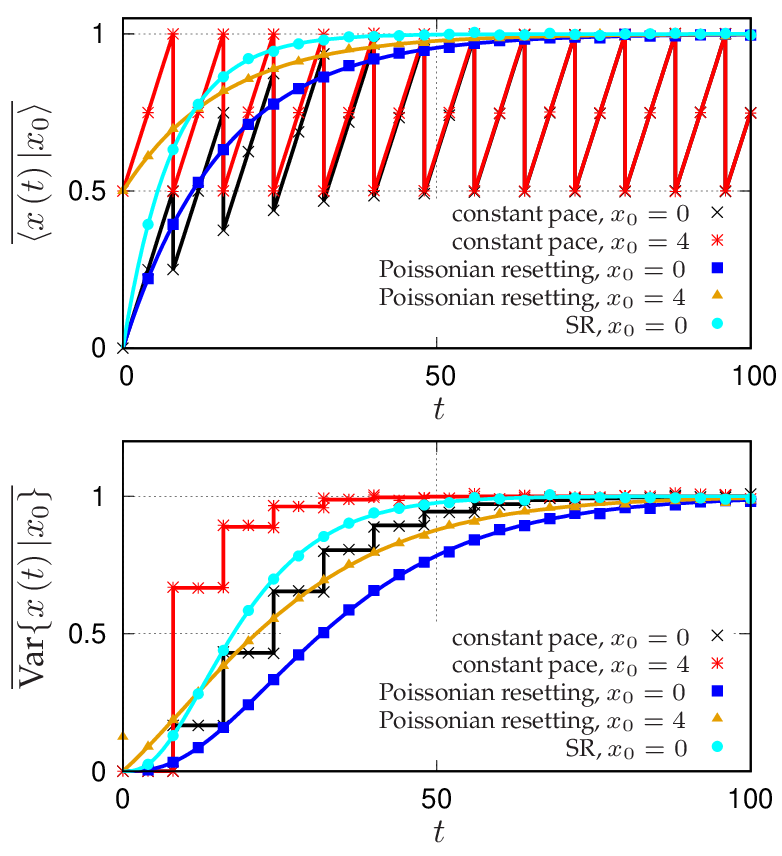}
\caption{Mean and variance of the height profile for dependent stochastic
resetting with Poissonian [$r=0.125$, Eqs.~(\ref{app_classical_reset_average})
and (\ref{app_classical_reset_variance})] and constant pace
[Eqs.~(\ref{aver_bal_uni_const}) and (\ref{var_bal_uni_const})] resetting
times for uniform resetting amplitude and
two different initial heights $x_0$, in comparison with classical resetting (SR).
We plot both quantities according to the normalization in
Eqs.~(\ref{app_dep_bal_aver_norm}) and (\ref{app_dep_bal_var_norm}).
The propagating process is ballistic ($v=0.5$) in all cases. Numerical results
are shown by points, the analytical results by solid lines.}
\label{average_and_variance_dep_reset}
\end{figure}

For Poissonian resetting times both mean and variance become independent of the
initial height in the long time limit. The functional behavior of both quantities
for Poissonian and constant pace resetting times are shown in
Fig.~\ref{average_and_variance_dep_reset}, in which we use the
nomalized expressions
\begin{eqnarray}
\overline{\langle x(t)|x_0\rangle}&=&\frac{\langle x(t)|x_0\rangle}{\underset{
t\to \infty}\limsup\langle x(t)|x_0\rangle},
\label{app_dep_bal_aver_norm}\\
\overline{\mathrm{Var}\{ x(t)|x_0\}}&=&\frac{\mathrm{Var}\{ x(t)|x_0\}}{\underset{
t\to\infty}\lim\mathrm{Var}\{ x(t)|x_0\}}.
\label{app_dep_bal_var_norm}
\end{eqnarray}
In this asymptotic limit the normalized mean converges to unity for Poissonian
resetting. In contrast, with constant pace resetting times the oscillating
quantity $\overline{\langle x(t)|x0\rangle}$ is limited from above by unity.
Based on definition (\ref{app_dep_bal_aver_norm}) of the normalized mean, the
two different convergence behaviors are compared in the upper panel of
Fig.~\ref{average_and_variance_dep_reset}. The normalized variance in
Eq.~(\ref{app_dep_bal_var_norm}) has the same limiting value for both
Poissonian and constant pace resetting, see the lower panel of
Fig.~\ref{average_and_variance_dep_reset}.

\subsection{Derivation of the probability density $P(x,t)$ for Poissonian resetting,
ballistic propagation and dependent resetting amplitudes}
\label{exp_res_int_length_another_approach_dep}

To derive a differential equation for the PDF $P(x,t;x_0;t_0)$ we use the fact
that the process is homogeneous in time and derived the master equation for $P(x,
t;x_0)$,
for which $(x(t+\Delta t)|x_0)=c(x(t)|x_0)$ with probability $r\Delta t$ and $(x(
t+\Delta t)|x_0)=x(t)|x_0)+v\Delta t$ with probability $1-r\Delta t$,
\begin{equation}
\frac{\partial P(x,t;x_0)}{\partial t}=-v\frac{\partial P(x,t;x_0)}{
\partial x}-r P(x,t;x_0)+r\int\limits_0^\infty\frac{dy}{y}P(y,t;x_0)f_C\left(
\frac{x}{y}\right),
\end{equation}
with $P(x,0;x_0)=\delta(x-x_0)$.
For the Laplace transform $\bar{P}(u,t;x_0)$ of $P(x,t;x_0)$ with respect to $x$
this yields
\begin{equation}
\label{another_approach_dep_char}
\frac{\partial\bar{P}(u,t;x_0)}{\partial t}=-uv\bar{P}(u,t;x_0)-r\bar{P}(u,t;x_0)
+r\int\limits_0^1dc\bar{P}(uc,t;x_0)f_C(c)
\end{equation}
with $\bar{P}(u,0;x_0)=\exp(-ux_0)$.

\subsubsection{Comparison with classical Stochastic Resetting}
\label{class_res}

If we assume a standard SR to the initial condition $x_0$ we have $f_C(c)=\delta(
c)$.  Moreover, the relation of the corresponding random variable, and thus the
partial differential is slightly different. Explicitly, $(x(t+\Delta t)|x_0)=
c\times(x(t)|x_0)+x_0$ with probability $r\Delta t$ and $(x(t+\Delta t)|x_0)=
(x(t)|x_0)+v\Delta t$ with probability $1-r\Delta t$, thus
\begin{eqnarray}
\nonumber
\frac{\partial P(x,t;x_0)}{\partial t}&=&-v\frac{\partial P(x,t;x_0)}
{\partial x}-rP(x,t;x_0)+r \int_0^\infty \frac{P(y,t;x_0)}{y} \delta
\left(\frac{x-x_0}{y}\right) dy\\
&=&\frac{\partial P(x,t;x_0)}{\partial t}=-v\frac{\partial P(x,t;x_0)}{\partial x}
-rP(x,t;x_0)+r \delta(x-x_0)
\label{another_approach_SR}
\end{eqnarray}
where $P(x,0;x_0)=\delta(x-x_0)$ and
we used the condition that $P(x,t;x_0)$ is normalized and the scaling
property of the delta function, $\delta(ax)=\delta(x)/|a|$ for $a\in\mathbb{R}$.
Finally, in the case of SR with an arbitrary initial distribution $\phi_0(x)$ the 
distribution of $x$ at time $t$ can be computed from $\rho(x,t)=\int_0^\infty 
\phi_0(x_0) P(x,t;x_0) dx_0$ and we get
\begin{equation}
\label{another_approach_CR2}
\frac{\partial \rho(x,t)}{\partial t}=-v\frac{\partial \rho(x,t)}{\partial x}-r
\rho(x,t)+r \phi_0(x)
\end{equation}
with $\rho(x,0)=\phi_0(x)$.
Eq.~\eqref{another_approach_SR} is homogeneous in space and confirms the 
results of Ref.~\cite{Evans1} for ballistic displacement instead of a diffusive
displacement.

\subsubsection{Stationary distribution for ballistic displacement, uniform
dependent resetting amplitude and Poissonian resetting}

We get the stationary solution of Eq.~(\ref{another_approach_dep_char})
for $f_C(c)=1$ with $P^*(x)=\underset{t\to\infty}{\lim} P(x,t;x_0)$ for
$\lim_{t\to\infty}\partial P(x,t;x_0)/\partial t=0$. Thus, for the spatial
Laplace transform $\bar{P}^*(u)$ becomes
\begin{eqnarray}
0=-uv\bar{P}^*(u)-r\bar{P}^*(u)+r\int_0^1\bar{P}^*(uc)dc,\quad
\Leftrightarrow\quad u(uv+r)\bar{P}^*(u)=r\int_0^u\bar{P}^*(c')dc',
\label{stat_another_dep_lap_intermittent}
\end{eqnarray}
with $c'=uc$.
If we now differentiate Eq.~\eqref{stat_another_dep_lap_intermittent} with 
respect to $u$ and use the normalization condition $\bar P^*(0)=1$, we get
\begin{equation}
(2uv+r)\bar{P}^*(u)+u(uv+r)\bar{P}^{*'}(u)=r\bar{P}^*(u),
\end{equation}
implying $\bar{P}^{*'}(u)=\frac{2v}{uv+r}\bar{P}^*(u)$ and $\bar{P}^*(0)=1$.
The solution is given by
\begin{equation}
\bar{P}^*(u)= \frac{r^2}{(uv+r)^2}.
\label{sol_stat_another_dep_lap}
\end{equation}
Eq.~\eqref{sol_stat_another_dep_lap} solves
Eq.~\eqref{stat_another_dep_lap_intermittent}, which proves our claim.

Thus, the stationary solution $P^*(x)$ is the inverse Laplace transform of 
$\bar{P}^{*}(u)$, Eq.~\eqref{sol_stat_another_dep_lap},
\begin{eqnarray}
P^*(x)=\lim_{t\to\infty} P(x,t;x_0)=\left(\frac{r}{v}\right)^2 x\exp\left(-
\frac{rx}{v}\right).
\label{sol_stat_another_dep}
\end{eqnarray}

\subsubsection{Proof of equality between partial differential equation
(\ref{another_approach_dep_char}) and integral representation
(\ref{app_dep_bal_lap_lap})}
\label{eq_intdiff}

Let $\tilde{\bar{P}}(u,s)$ denote the Laplace transform of $\bar{P}(u,t)$, we can
obtain the following forms for Poissonian resetting in double-Laplace space,
\begin{eqnarray}
\label{another_approach_dep_char_Lap}
\tilde{\bar{P}}(u,s;x_0)=\frac{\exp(-ux_0)}{r+s+uv}+\frac{r}{r+s+uv}\int
\limits_0^1dc\tilde{\bar{P}}(uc,s;x_0)f_C(c)
\end{eqnarray}
with the following iterative approximations
\begin{eqnarray}
\nonumber
&&0\mbox{th approximation}=\frac{\exp(-ux_0)}{r+s+uv}\\
\nonumber
&&1\mbox{st approximation}=\frac{\exp(-ux_0)}{r+s+uv}+\frac{r}{r+s+uv}\int
\limits_0^1dc\frac{\exp(-ux_0c)}{r+s+uvc})f_C(c)\\
\nonumber
&&2\mbox{nd approximation}=\frac{\exp(-ux_0)}{r+s+uv}+\frac{r}{r+s+uv}\int
\limits_0^1dc_1\frac{\exp(-ux_0c_1)}{r+s+uvc_1}f_C(c_1)\\
\nonumber
&&+\frac{r}{r+s+uv}\int\limits_0^1dc_1\frac{rf_C(c_1)}{r+s+uvc_1}\int\limits_0^1
dc_2\frac{\exp(-ux_0c_1c_2)f_C(c_2)}{r+s+uvc_1 c_2}\\
\nonumber
&&n\mbox{th approximation}=\frac{\exp(-ux_0)}{r+s+uv}+\frac{1}{r+s+uv}\sum^n_{m=1}
\left(\prod^m_{j=1}\int\limits_0^1dc_j\frac{r f_C(c_j)}{\left(r+s+uv\prod^j_{
i=1}c_i\right)}\right)\exp\left(-x_0\prod^m_{j=1}c_j\right)
\end{eqnarray}
such that we find
\begin{eqnarray}
\tilde{\bar{P}}(u,s;x_0)=\frac{\exp(-ux_0)}{r+s+uv}+\frac{1}{r+s+uv}\sum^{\infty}
_{n=1}\left(\prod^n_{j=1}\int\limits_0^1dc_j\frac{rf_C(c_j)}{\left(r+s+uv\prod^j
_{i=1}c_i\right)}\right)\exp\left(-x_0\prod^n_{j=1}c_j\right),
\label{another_approach_dep_char_sol_explicit}
\end{eqnarray}
which is equal to Eq.~(\ref{app_dep_bal_lap_lap}) for Poissonian resetting, and
thus proves our claim.
\end{widetext}

\subsection{Graphical illustration for dependent resetting}

We finally illustrate the difference between ballistic propagation with Poissonian
and constant pace resetting for uniform dependent resetting amplitude. To this end
we compare the corresponding PDFs at different times respectively show the behavior
of mean and variance of $(x(t)|x_0)$.

\begin{figure*}
\includegraphics{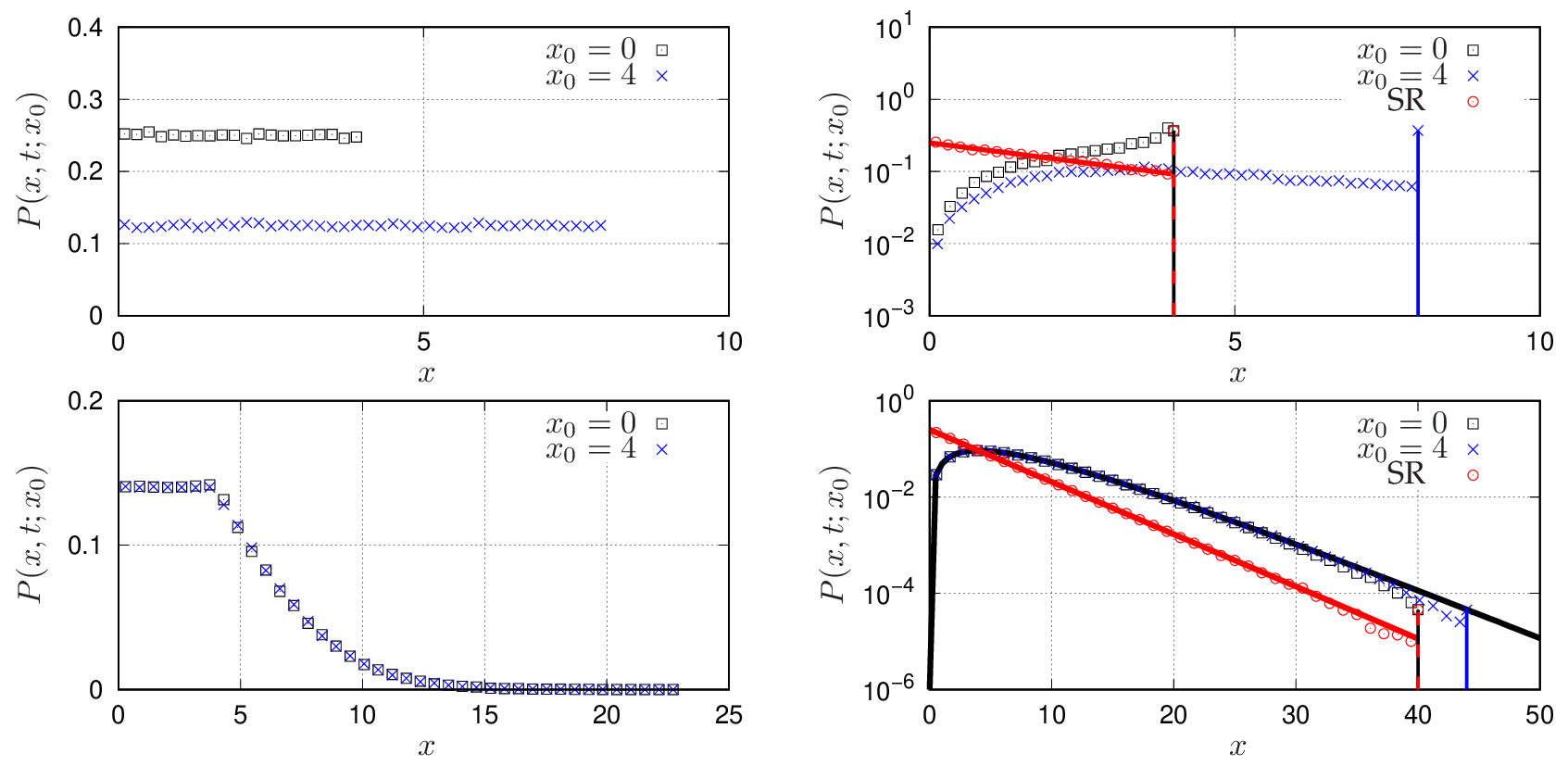}
\vspace*{-0.3cm}
\caption{PDF $P(x,t;x_0)$ of the height profile for different initial heights 
and ballistic motion with uniform resetting. Left: "constant pace" resetting.
Right: Poissonian resetting, compared to the classical resetting scenario with
enforced resets to the origin. Top: $t=1/r$. Bottom: $t=10/r$. Numerical
results are shown by points, analytical results by solid lines. Parameters:
$v=0.5$ and $r=0.125$.}
\label{app_dep_pdf}
\end{figure*}

Fig.~\ref{app_dep_pdf} shows the position PDF for ballistic displacement,
uniformly distributed resetting amplitude and two different distributions
of resetting interval lengths. For each process the impact of different
initial values $x_0$ is shown. It is obvious that the influence of initial
values eventually disappears, as can be seen in the upper panels. In the
left panel of Fig.~\ref{app_dep_pdf} constant pace resetting is used. When 
the impact of the initial value disappears (lower left panel) the PDF of $x$ has a 
uniform part for small values of $x$. However, the uniform character disappears 
from a certain value of $x$ and decreases in the tail. The distribution does not 
change its shape, however, the PDF of $x$ fulfills a periodic movement. This motion 
of the distribution $P(x,t;x_0)$ is divided in a linear shift in time and a shift 
in the opposite direction as a point process in time. In the right panels of
Fig.~\ref{app_dep_pdf} Poissonian resetting is used. The height of the probability 
of no resets is independent of the value of $x_0$. This probability is mapped at $x=
vt+x_0$ and decreasing in time. For longer $t$ (right lower panel) it can be 
seen that the process is stationary.

\begin{figure*}
\includegraphics{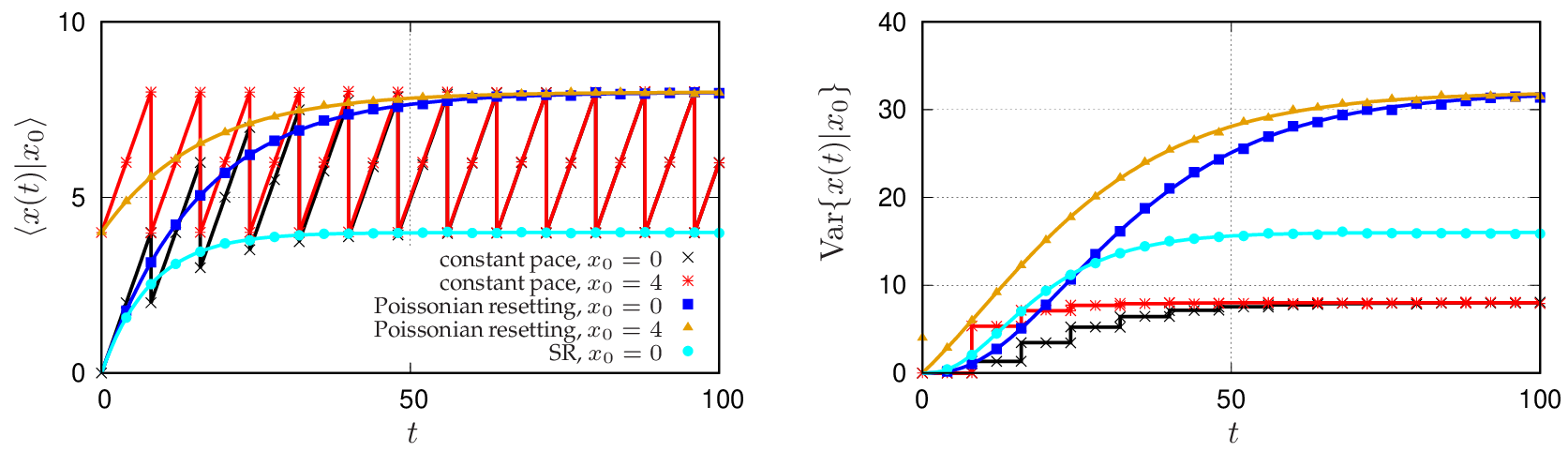}
\caption{Mean $\langle x(t)|x_0\rangle$ and variance $\mathrm{Var}\{ x(t)|x_0\}$ of 
the height profile for dependent stochastic resetting with Poissonian and 
"constant pace" resetting times for uniform resetting amplitude and two different 
initial heights $x_0$, in comparison with classical resetting. The 
propagating process is ballistic ($v=0.5$) in all cases. For both types of resetting
the resetting rate is $r=0.125$. Numerical results are
shown by points, the analytical results by lines.}
\label{app_average_and_variance_dep_reset}
\end{figure*}

In Fig.~\ref{app_average_and_variance_dep_reset}  we can see the temporal behavior
of the average and variance of $(x(t)|x_0)$. We show the results for ballistic
displacement process which is interrupted by uniform dependent resetting events
for two different distributions of resetting interval lengths. All analytical
results are numerically verified, see Fig.~\ref{app_average_and_variance_dep_reset}.
The vanishing impact of different initial values $x_0$ for average and variance of
$(x(t)|x_0)$ with $t$ can be seen in all panels. The average $\langle x(t)|x_0
\rangle$ (left upper panel) increases linearly with $t$ during the constant
resetting interval lengths and decreases at the resetting points. After some time
the average of $(x(t)|x_0)$ is confined to a certain range and has a periodic
switch between linear increase and decrease as a point process in time. The 
corresponding $\mathrm{Var}\{x(t)|x_0\}$ (left lower panel) stays the same during 
the resetting interval lengths and increases discontinuously at the resetting points, 
a jump in the figure. For longer $t$ the variance $\mathrm{Var}\{x(t)|x_0\}$ 
converges to a finite limit. In the right panels of
Fig.~\ref{app_average_and_variance_dep_reset} the convergence of average and
variance of $(x(t)|x_0)$ in presence of Poissonian resetting is obvious. Thus,
this process is stationary.

\begin{figure}
\includegraphics[width=6.8cm]{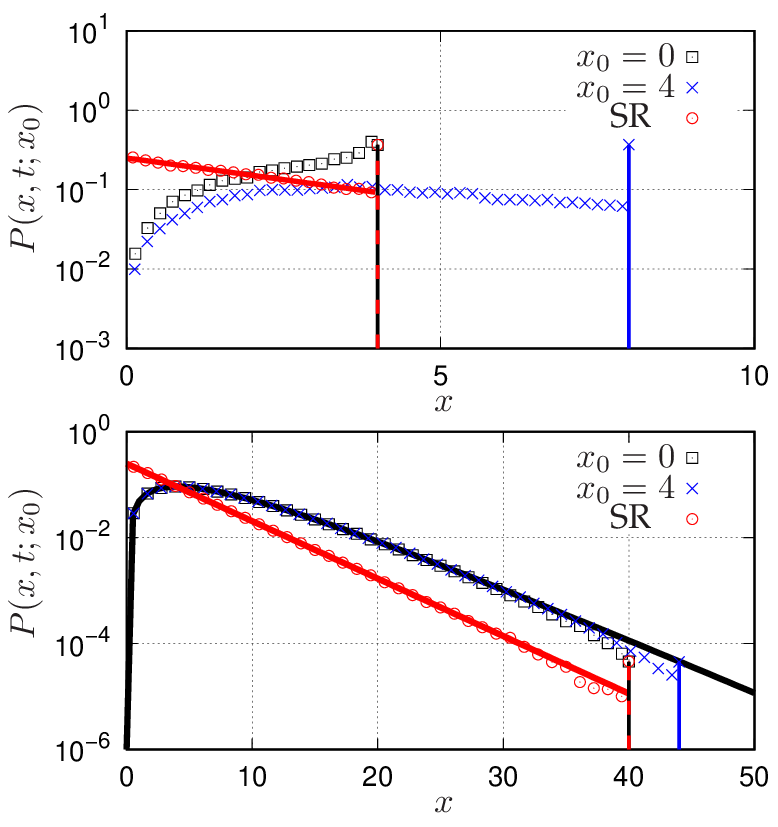}
\caption{PDF $P(x,t;x_0)$ of the height profile for different initial heights and
uniform resetting amplitude, compared to the classical resetting scenario with
enforced resets to the origin, for ballistic propagation ($v=0.5$) and Poissonian
resetting times ($r=0.125$). Top: $t=1/r$; bottom: $t=10/r$. Numerical results
are shown by points, analytical results by lines.}
\label{dep_pdf}
\end{figure}

Fig.~\ref{dep_pdf} shows the PDF for ballistic propagation with Poissonian
resetting times for classical resetting to the origin and uniform resetting
amplitudes, for two different initial heights
$x_0$. At early times of the process (top panel) the difference due to the
initial height is distinct, while in the long time limit (bottom panel) the
PDFs for the two uniform-resetting cases coincide. The difference to the
classical resetting case with enforced resetting to the origin clearly results
in a lower height profile.

\section{Conclusions}
\label{conc}

We introduced a generalized resetting concept with random resetting amplitudes
in two different scenarios: independent resetting, in which the height profile
may become negative, depending on the specific resetting amplitude PDF and the
propagating process; and dependent resetting, in which the positivity of the
height profile is guaranteed by the definition of the resetting amplitude
PDF. We derived an explicit analytical formulation of the process and analyzed
specifically ballistic propagation in the presence of Poissonian resetting times
and different resetting amplitude PDFs. We also demonstrated that the
classical resetting theory with mandatory resets to the origin is contained in
our model in the dependent case, whereas the independent scenario is a specific
case of jump diffusion \cite{Kou} with one-sided jump lengths.

Physically, the RASR process introduce here corresponds to the scenario of a
propagating stochastic or deterministic process, that is interrupted by random
resets. This may correspond to the geophysical stratigraphic scenario, in which
the propagation mimics the gradual build-up or decay of a sedimentation profile,
whereas the resets represent sudden erosion events. The latter could be seasonal
("constant pace") or random-in-time weather events such as extreme floods. In fact
our model is similar (albeit more flexible) to that proposed in \cite{SD_6}, where
constant rates of accumulation were considered the null hypothesis and the effect
of random erosion periods on bed hiatus length distributions were explored. We
also note similar strategies developed for ecohydrology applications \cite{iturbe},
and the general development of a class of jump processes \cite{porporato}. In a
different context we could think of population dynamics interrupted by epidemics,
pathogens (e.g., embodied by bacterial biofilms) decimated by antibiotic
treatment (here, both periodic and "random" application protocols are being
employed in clinical studies), or crises-interrupted financial markets. All
these cases process correspond to the intermittent picture of a parent
process (the "propagation") with superimposed resetting statistic.

The qualitative difference between independent and dependent resetting is that
the latter case becomes stationary for ballistic propagation and Poissonian
resetting times,
whereas the former remains non-stationary. The fact that our basic model can be
recast in these two variants underlines the flexibility embedded in this simple
extension of classical resetting (SR). Another appeal is the relatively
straightforward, fully analytical description, with the caveat that not all
resulting expressions can be expressed fully explicitly. Having said this, we
believe that our results represent an attractive extension of the resetting
process. Apart from the above physical scenarios the described flexibility of
our extension of the resetting dynamics will be of interest in the mathematical
theory of random search processes.

\begin{acknowledgments}
This research is supported by the Basque Government through the BERC 2018-2021 
program and by the Ministry of Science, Innovation and Universities: BCAM Severo 
Ochoa accreditation SEV-2017-0718. We acknowledge support from DFG (ME 1535/7-1). 
RM acknowledges the Foundation for Polish Science (Fundacja na rzecz Nauki Polskiej,
FNP) for support within an Alexander von Humboldt Honorary Polish Research
Scholarship.
\end{acknowledgments}

\clearpage

\appendix

\onecolumngrid

\section{Mathematical identity between first and last resetting picture}
\label{identity_equal}

In this section we prove the following formal mathematical identity that will
be used in Appendix B below to demonstrate the equivalence of the first and the
last resetting pictures:
\begin{eqnarray}
\nonumber
&&\prod^n_{j=1}\left(\int_{t_{j-1}}^tdt_j\int_{A_1}^{A_2}dy_j\int_{A_3}^{A_4}dz_j
\eta_1(t_j,t_{j-1},y_j,y_{j-1},z_j,z_{j-1})\right)\eta_2(x,t,t_n,y_n,z_n)\\
\nonumber
&&=\int_0^{t-t_0}d\tau_n\int_{A_1}^{A_2}dy_n\int_{A_3}^{A_4}dz_n\left(\prod^{n-1}
_{i=1}\int_0^{\tau_{n+1-i}}d\tau_{n-i}\int_{A_1}^{A_2}dy_{n-i}\int_{A_3}^{A_4}dz
_{n-i}\right.\\
\nonumber
&&\times\eta_1(\tau_{n+1-i}+t_0,\tau_{n-i}+t_0,y_{n+1-i},y_{n-i},z_{n+1-i},z_{n-i})
\Big)\\
\nonumber
&&\times\eta_1(\tau_1+t_0,t_0,y_1,y_0,z_1,z_0)\eta_2(x,t,\tau_n +t_0,y_n,z_n)
\end{eqnarray}

\begin{eqnarray}
\nonumber
&&\Leftrightarrow\prod^n_{j=1}\left(\int_{\tau_{j-1}}^{t-t_0}d\tau_j\int^{A_2}
_{A_1}dy_j\int^{A_4}_{A_3}dz_j\eta_1(\tau_j+t_0,\tau_{j-1}+t_0,y_j,y_{j-1},z_j,
z_{j-1})\right)\eta_2(x,t,\tau_n+t_0,y_n,z_n)\\
\nonumber
&&=\int_0^{t-t_0}d\tau_n\int_{A_1}^{A_2}dy_n\int_{A_3}^{A_4}dz_n\left(\prod
^{n-1}_{i=1}\int_0^{\tau_{n+1-i}}d\tau_{n-i}\int_{A_1}^{A_2}dy_{n-i}\int_{A_3}
^{A_4}dz_{n-i}\right.\\
\nonumber
&&\times\eta_1(\tau_{n+1-i}+t_0,\tau_{n-i}+t_0,y_{n+1-i},y_{n-i},z_{n+1-i},
z_{n-i})\Big)\\
&&\times\eta_1(\tau_1+t_0,t_0,y_1,y_0,z_1,z_0)\eta_2(x,t,\tau_n+t_0,y_n,z_n)
\label{identity}
\end{eqnarray}
with $\tau_j=t_j-t_0$ for $0\le j\le n$.
To prove Eq.~\eqref{identity} we use the method of induction. For $n=1$, the
Eq.~\eqref{identity} is obviously fulfilled,
\begin{eqnarray}
\nonumber
&&\int_0^{t-t_0}d\tau_1\int_{A_1}^{A_2}dy_1\int_{A_3}^{A_4}dz_1\eta_1(\tau_1
+t_0,t_0,y_1,y_0,z_1,z_0)\eta_2(x,t,\tau_1+t_0,y_1,z_1)\\
&&=\int_0^{t-t_0}d\tau_1\int_{A_1}^{A_2}dy_1\int_{A_3}^{A_4}dz_1\eta_1(\tau_1
+t_0,t_0,y_1,y_0,z_1,z_0)\eta_2(x,t,\tau_1+t_0,y_1,z_1).
\end{eqnarray}

Next, we take the inductive step $n\Rightarrow(n+1)$,
\begin{eqnarray}
\nonumber
&&\prod^{n+1}_{j=1}\left(\int\limits_{\tau_{j-1}}^{t-t_0}d\tau_j\int\limits
_{A_1}^{A_2}dy_j\int\limits_{A_3}^{A_4}dz_j\eta_1(\tau_j+t_0,\tau_{j-1}+t_0,
y_j,y_{j-1},z_j,z_{j-1})\right)\eta_2(x,t,\tau_{n+1}+t_0,y_{n+1},z_{n+1})\\
\nonumber
&&=\int\limits_{\tau_n}^{t-t_0}d\tau_{n+1}\int\limits_{A_1}^{A_2}dy_{n+1}
\int\limits_{A_3}^{A_4}dz_{n+1}\eta_1(\tau_{n+1}+t_0,\tau_n+t_0,y_{n+1},y_n,
z_{n+1},z_n)\\
\nonumber
&&\times\prod^n_{j=1}\left(\int\limits_{\tau_{j-1}}^{t-t_0}d\tau_j\int\limits
_{A_1}^{A_2}dy_j\int\limits_{A_3}^{A_4}dz_j\eta_1(\tau_j +t_0,\tau_{j-1} +t_0,
y_j,y_{j-1},z_j,z_{j-1})\right)\eta_2(x,t,\tau_{n+1}+t_0,y_{n+1},z_{n+1}),
\end{eqnarray}
i.e.,
\begin{eqnarray}
\nonumber
&&\int\limits_0^{t-t_0}d\tau_n\int\limits_{\tau_n}^{
t-t_0}d\tau_{n+1}\int\limits_{A_1}^{A_2}dy_n\int\limits_{A_3}^{A_4}dz_n\\
\nonumber
&&\times\left(\prod^{n-1}_{i=1}\int\limits_0^{\tau_{n+1-i}}d\tau_{n-i}\int
\limits_{A_1}^{A_2}dy_{n-i}\int\limits_{A_3}^{A_4}dz_{n-i}\eta_1(\tau_{n+1-i}
+t_0,\tau_{n-i}+t_0,y_{n+1-i},y_{n-i},z_{n+1-i},z_{n-i})\right)\\
\nonumber
&&\times\int\limits_{A_1}^{A_2}dy_{n+1}\int\limits_{A_3}^{A_4}dz_{n+1}\eta_1
(\tau_{n+1}+t_0,\tau_n+t_0,y_{n+1},y_n,z_{n+1},z_n)\eta_1(\tau_1+t_0,t_0,y_1,
y_0,z_1,z_0)\eta_2(x,t,\tau_n+t_0,y_n,z_n)\\
\nonumber
&&=\int\limits_0^{t-t_0}d\tau_{n+1}\int\limits_0^{\tau_{n+1}}d\tau_n\int
\limits_{A_1}^{A_2}dy_n\int\limits_{A_3}^{A_4}dz_n\\
\nonumber
&&\times\left(\prod^n_{i=2}\int\limits_0^{\tau_{n+2-i}}d\tau_{n+1-i}\int
\limits_{A_1}^{A_2}dy_{n+1-i}\int\limits_{A_3}^{A_4} dz_{n+1-i}\eta_1(\tau_
{n+2-i}+t_0,\tau_{n+1-i}+t_0,y_{n+2-i},y_{n+1-i},z_{n+2-i},z_{n+1-i})\right)\\
\nonumber
&&\times\int\limits_{A_1}^{A_2}dy_{n+1}\int\limits_{A_3}^{A_4}dz_{n+1}\eta_1(
\tau_{n+1}+t_0,\tau_n+t_0,y_{n+1},y_n,z_{n+1},z_n)\eta_1(\tau_1+t_0,t_0,y_1,y_0,
z_1,z_0)\eta_2(x,t,\tau_n+t_0,y_n,z_n)
\end{eqnarray}

\begin{eqnarray}
\nonumber
&&=\int\limits_0^{t-t_0}d\tau_{n+1}\int\limits_{A_1}^{A_2}dy_{n+1}\int\limits
_{A_3}^{A_4}dz_{n+1}\\
\nonumber
&&\times\left(\prod^{n}_{i=1}\int\limits_0^{\tau_{n+2-i}} d\tau_{n+1-i}\int
\limits_{A_1}^{A_2}dy_{n+1-i}\int\limits_{A_3}^{A_4}dz_{n+1-i}\eta_1(\tau_{
n+2-i}+t_0,\tau_{n+1-i}+t_0,y_{n+2-i},y_{n+1-i},z_{n+2-i},z_{n+1-i})\right)\\
&&\times\eta_1(\tau_1+t_0,t_0,y_1,y_0,z_1,z_0)\eta_2(x,t,\tau_n+t_0,y_n,z_n).
\end{eqnarray}
This proves our claim.

\section{Derivation of last resetting picture for independent resetting amplitudes}
\label{indep_equal}

In this section we aim to show the equivalence of the description in the
\emph{first resetting picture\/},
\begin{eqnarray}
\label{p_r_indep_first_app}
P(x,t;x_0,t_0)=\Psi(t-t_0)G(x,t;x_0,t_0)+\int\limits_{t_0}^t dt_1\psi(t_1-t_0)
\int\limits_{-\infty}^\infty dy G(y,t_1;x_0,t_0)\int\limits_{-\infty}^\infty
dx_1q\left(x_1-y\right)P(x,t;x_1,t_1)
\end{eqnarray}
and the \emph{last resetting picture\/} that includes all resetting steps,
\begin{eqnarray}
\nonumber
&&P(x,t;x_0,t_0)=\Psi(t-t_0)G(x,t;x_0,t_0)+\sum^\infty_{n=1}\int\limits_0^{t-t_0}
d\tau_n\int\limits_{-\infty}^\infty dx_n\int\limits_{-\infty}^\infty dy_n\\
\nonumber
&&\times\left(\prod^{n-1}_{i=1}\int\limits_0^{\tau_{n+1-i}} d\tau_{n-i}\psi(\tau_{
n+1-i}-\tau_{n-i})\int\limits_{-\infty}^\infty dy_{n-i}G(y_{n+1-i},\tau_{n+1-i}+t_0
;x_{n-i},\tau_{n-i}+t_0)\int\limits_{-\infty}^\infty dx_{n-i}q\left(x_{n+1-i}-y_{
n+1-i}\right)\right)\\
&&\times q\left(x_1-y_1\right)\psi(\tau_1)G(y_1,\tau_1+t_0;x_0,t_0)\Psi(t-t_0
-\tau_n)G(x,t;x_n,\tau_n +t_0).
\label{p_r_indep_last_app}
\end{eqnarray}
To this end we write Eq.~\eqref{p_r_indep_last_app} as
\begin{eqnarray}
\nonumber
&&P(x,t;x',t')=\Psi(t-t')G(x,t;x',t')+\\
&&+\sum^\infty_{n=1}\left(\prod^n_{j=1}\int\limits_{t_{j-1}}^tdt_{j}\psi(t_j-
t_{j-1})\int\limits_{-\infty}^\infty dy_jG(y_j,t_j ;x_{j-1},t_{j-1})\int
\limits_{-\infty}^\infty dx_jq\left(x_j-y_j\right)\right)\Psi(t-t_n)G(x,t;x_n,
t_n)
\label{p_r_indep_last_t_app}
\end{eqnarray}
with $t_0=t'$ and $x_0=x'$. The equivalence of Eqs.\eqref{p_r_indep_last_app} and
\eqref{p_r_indep_last_t_app} will be proven in this section below. Now we
substitute $P(x,t;x_0,t_0)$ and $P(x,t;x_1,t_1)$ in the first resetting
picture, Eq.~\eqref{p_r_indep_first_app} with Eq.~\eqref{p_r_indep_last_t_app}.
The left hand side (LHS) of Eq.~\eqref{p_r_indep_first_app} after this
substitution becomes
\begin{eqnarray}
\nonumber
&&\text{LHS}=P(x,t;x_0,t_0)=\Psi(t-t_0)G(x,t;x_0, t_0)+\\
&&+\sum^\infty_{n=1}\left(\prod^n_{j=1}\int\limits_{t_{j-1}}^tdt_j\psi(t_j
-t_{j-1})\int\limits_{-\infty}^\infty dy_jG(y_j,t_j;x_{j-1},t_{j-1})\int
\limits_{-\infty}^\infty dx_jq\left(x_j-y_j\right)\right)\Psi(t-t_n)G(x,t;
x_n,t_n).
\label{p_r_indep_last_t_LHS_app}
\end{eqnarray}
As $P(x,t;x_1,t_1)$ in Eq.~\eqref{p_r_indep_first_app} has the initial value
$x_1$ at $t_1$, these two variables have the lowest index $1$ instead of $0$,
and thus instead of Eq.~\eqref{p_r_indep_last_t_LHS_app} one gets
\begin{eqnarray}
\nonumber
&&P(x,t;x_1,t_1)=\Psi(t-t_1)G(x,t;x_1,t_1)\\
&&+\sum^\infty_{n=2}\left(\prod^n_{j=2}\int\limits_{t_{j-1}}^tdt_j\psi(t_j-
t_{j-1})\int\limits_{-\infty}^\infty dy_j G(y_j,t_j;x_{j-1},t_{j-1})\int
\limits_{-\infty}^\infty dx_jq(x_j-y_j)\right)\Psi(t-t_n)G(x,t;x_n,t_n).
\label{p_r_indep_last_app_1}
\end{eqnarray}
Substituting \eqref{p_r_indep_last_app_1} into the RHS of
Eq.~\eqref{p_r_indep_first_app} we get
\begin{eqnarray}
\nonumber
&&\mathrm{RHS}=\Psi(t-t_0)G(x,t;x_0, t_0)\\
\nonumber
&&+\int\limits_{t_0}^tdt_1\psi(t_1-t_0)\int\limits_{-\infty}^\infty dyG(y,t_1;
x_0,t_0)\int\limits_{-\infty}^\infty dx_1q(x_1-y)\Psi(t-t_1)G(x,t;x_1,t_1)\\
\nonumber
&&+\int\limits_{t_0}^tdt_1\int\limits_{-\infty}^\infty dy\int\limits_{-\infty}
^\infty dx_1\sum^\infty_{n=2}\left(\prod^n_{j=2}\int\limits_{t_{j-1}}^tdt_j
\psi(t_j-t_{j-1})\int\limits_{-\infty}^\infty dy_jG(y_j,t_j;x_{j-1},t_{j-1})
\int\limits_{-\infty}^\infty dx_jq\left(x_j-y_j\right)\right)\\
\nonumber
&&\times\psi(t_1-t_0)G(y,t_1;x_0,t_0)q(x_1-y)\Psi(t-t_n)G(x,t;x_n,t_n)\\
\nonumber
=&&\Psi(t-t_0)G(x,t;x_0,t_0)\\
\nonumber
&&+\int\limits_{t_0}^tdt_1\psi(t_1-t_0)\int\limits_{-\infty}^\infty dy_1G(y_1,
t_1;x_0,t_0)\int\limits_{-\infty}^\infty dx_1q(x_1-y_1)\Psi(t-t_1)G(x,t;x_1,t_1)\\
&&+\sum^\infty_{n=2}\left(\prod^n_{j=1}\int\limits_{t_{j-1}}^tdt_j\psi(t_j-t_{j-1})
\int\limits_{-\infty}^\infty dy_jG(y_j,t_j;x_{j-1},t_{j-1})\int\limits_{-\infty}
^\infty dx_jq(x_j-y_j)\right)\Psi(t-t_n)G(x,t;x_n,t_n)
\end{eqnarray}
with $y_1=y$. Then,
\begin{eqnarray}
\nonumber
&&\mathrm{RHS}=\Psi(t-t_0)G(x,t;x_0, t_0)\\
+&&\sum^\infty_{n=1}\left(\prod^n_{j=1}\int\limits_{t_{j-1}}^tdt_j\psi(t_j-t_{
j-1})\int\limits_{-\infty}^\infty dy_jG(y_j,t_j;x_{j-1},t_{j-1})\int\limits_{
-\infty}^\infty dx_jq(x_j-y_j)\right)\Psi(t-t_n)G(x,t;x_n,t_n),
\label{p_r_indep_last_t_RHS_app}
\end{eqnarray}
and thus $\mathrm{RHS}=\mathrm{LHS}$, which proves our claim. Thus,
Eq.~\eqref{p_r_indep_last_t_app} solves the \emph{first resetting picture\/}
of Eq.~\eqref{p_r_indep_first_app}, and Eq.~\eqref{p_r_indep_last_t_app}
describes the RASR with independent resetting amplitudes. If we can show that
Eq.~\eqref{p_r_indep_last_t_app} and the \emph{last resetting picture\/} of
Eq.~\eqref{p_r_indep_last_app} are equal, we demonstrate that both mathematical
representations describe the same process. To this end, consider
\begin{eqnarray}
\nonumber
&&\mathrm{LHS}=P(x,t;x_0,t_0)=\Psi(t-t_0)G(x,t;x_0,t_0)\\
\nonumber
&&+\sum^\infty_{n=1}\left(\prod^n_{j=1}\int\limits_{t_{j-1}}^tdt_j\psi(t_j-
t_{j-1})\int\limits_{-\infty}^\infty dy_jG(y_j,t_j;x_{j-1},t_{j-1})\int
\limits_{-\infty}^\infty dx_jq(x_j-y_j)\right)\Psi(t-t_n)G(x,t;x_n,t_n)\\
\nonumber
&&=\Psi(t-t_0)G(x,t;x_0,t_0)\\
\nonumber
&&+\sum^\infty_{n=1}\left(\prod^n_{j=1}\int\limits_{\tau_{j-1}}^{t-t_0}d\tau_j
\psi(\tau_j-\tau_{j-1})\int\limits_{-\infty}^\infty dy_jG(y_j,\tau_j+t_0
;x_{j-1},\tau_{j-1}+t_0)\int\limits_{-\infty}^\infty dx_jq(x_j-y_j)
\right)\\
&&\times\Psi(t-\tau_n)G(x,t;x_n,\tau_n +t_0),
\end{eqnarray}
with $\tau_j=t_j-t_0$ for $1\le j\le n$.

If we now use Eq.~\eqref{identity} with the substitution \eqref{indep_subs}, we
obtain
\begin{eqnarray}
\label{indep_subs}
&&\left\{\begin{array}{lllll}
\eta_1(t_j,t_{j-1},y_j,y_{j-1},z_j,z_{j-1})=\psi(t_j-t_{j-1})G(y_j,t_j;z_{j-1},
t_{j-1})q(z_j-y_j)\\[0.2cm]
\eta_2(x,t,t_n,y_n,z_n)=\Psi(t-t_0)G(x,t;z_n,t_n)\\[0.2cm]
z_j=x_j, \,\,\, t_j=\tau_j+t_0\\[0.2cm]
A_1,A_3=-\infty,\,\,\, A_2,A_4=\infty
\end{array}
\right.,
\end{eqnarray}
for $1\le j\le n$. We then find
\begin{eqnarray}
\nonumber
&&\text{LHS}=\Psi(t-t_0)G(x,t;x_0,t_0)+\sum^\infty_{n=1}\int\limits_0^{t-t_0}
d\tau_n\int\limits_{-\infty}^\infty dx_n\int\limits_{-\infty}^\infty dy_n
\left(\prod^{n-1}_{i=1}\int\limits_0^{\tau_{n+1-i}}d\tau_{n-i}\psi(\tau_{n+1
-i}-\tau_{n-i})\right)\\
\nonumber
&&\times\left(\prod^{n-1}_{i=1}\int\limits_{-\infty}^\infty dy_{n-i}G(y_{
n+1-i},\tau_{n+1-i}+t_0;x_{n-i},\tau_{n-i}+t_0)\int\limits_{-\infty}
^\infty dx_{n-i}q(x_{n+1-i}-y_{n+1-i})\right)\\
&&\times q(x_1-y_1)\psi(\tau_1)G(y_1,\tau_1+t_0;x_0,t_0)\Psi(t-t_0
-\tau_n)G(x,t;x_n,\tau_n +t_0),\label{indep_last_derive}
\end{eqnarray}
which represents exactly the \emph{last resetting picture} \eqref{p_r_indep_last_app}, 
proving our claim.

If we assume a free propagator, that is homogeneous in space and in time, the
stochastic process with resetting itself will be homogeneous in space and time,
$G(x,t;x_0,t_0)=G(x-x_0,t-t_0;0,0)\Rightarrow P(x,t;x_0,t_0)=P(x-x_0,t-t_0;0,0)$.
By assuming $G(x,t;x_0,t_0)=G(x-x_0,t-t_0;0,0)$ the density $P(x,t;x_0,t_0)$,
Eq.~\eqref{indep_last_derive}, then becomes
\begin{eqnarray}
\nonumber
&&P(x,t;x_0,t_0)=\Psi(t-t_0)G(x-x_0,t-t_0;0,0)+\sum^\infty_{n=1}\int\limits_0^{t-t_0}
d\tau_n\int\limits_{-\infty}^\infty dx_n\int\limits_{-\infty}^\infty dy_n\\
\nonumber
&&\times\left(\prod^{n-1}_{i=1}\int\limits_0^{\tau_{n+1-i}} d\tau_{n-i}\psi(\tau_{
n+1-i}-\tau_{n-i})\int\limits_{-\infty}^\infty dy_{n-i}G(y_{n+1-i}-x_{n-i},
\tau_{n+1-i}-\tau_{n-i};0,0)\int\limits_{-\infty}^\infty dx_{n-i}q\left(x_{n+1-i}-
y_{n+1-i}\right)\right)\\
\nonumber
&&\times q\left(x_1-y_1\right)\psi(\tau_1)G(y_1-x_0,\tau_1;0,0)\Psi(t-t_0
-\tau_n)G(x-x_n,t-t_0-\tau_n;0,0),\\
\nonumber
&&=\Psi(t-t_0)G(x-x_0,t-t_0;0,0)+\sum^\infty_{n=1}\int\limits_0^{t-t_0} d\tau_n\int
\limits_{-\infty}^\infty dx^{\prime}_n\int\limits_{-\infty}^\infty dy^{\prime}_n
\label{indep_homogen_app}\\
\nonumber
&&\times\left(\prod^{n-1}_{i=1}\int\limits_0^{\tau_{n+1-i}} d\tau_{n-i}\psi(\tau_{
n+1-i}-\tau_{n-i})\int\limits_{-\infty}^\infty dy^{\prime}_{n-i}G(y^{\prime}_{n+1-i}
-x^{\prime}_{n-i},\tau_{n+1-i}-\tau_{n-i};0,0)\int\limits_{-\infty}^\infty 
dx^{\prime}_{n-i}q\left(x^{\prime}_{n+1-i}-y^{\prime}_{n+1-i}\right)\right)\\
&&\times q\left(x^{\prime}_1-y^{\prime}_1\right)\psi(\tau_1)G(y^{\prime}_1,\tau_1
;0,0)\Psi(t-t_0-\tau_n)G(x-x_0-x^{\prime}_n,t-t_0-\tau_n;0,0),
\end{eqnarray}
in which $x^{\prime}_j=x_j-x_0$ and  $y^{\prime}_j=y_j-x_0$ for $1\le j\le n$.
On the right hand side of Eq.~\eqref{indep_homogen_app} $x$ and $x_0$ as well as
$t$ and $t_0$ only occur as differences $x-x_0$ and $t-t_0$ and not as single terms.
Thus, $G(x,t;x_0,t_0)=G(x-x_0,t-t_0;0,0)\Rightarrow P(x,t;x_0,t_0)=P(x-x_0,t-t_0;
0,0)$, which proves our claim.

\section{Differential equation for $P(x,t)$ with Poissonian resetting,
ballistic displacement process, and arbitrary independent resetting amplitudes}
\label{different}

To derive a differential equation for the PDF $P(x,t;x_0;t_0)$ we use the fact
that the process is homogeneous in space and time. We use the short-hand form
$P(x,t)$ for the choice $x(t_0=0)=0$. As the $x$-propagation for ballistic
motion reads
\begin{eqnarray}
x(t+\Delta t)=\left\{\begin{array}{ll}x(t)+z\mbox{ with probability }r\Delta t\\
x(t)+v\Delta t \mbox{ with probability }1-r\Delta t\end{array}\right..
\end{eqnarray}
This means that
\begin{eqnarray}
\label{another_approach_indep}
\left\{\begin{array}{ll}\displaystyle\frac{\partial P(x,t)}{\partial t}=-v
\frac{\partial P(x,t)}{\partial x}-rP(x,t)+r\int\limits_{-\infty}^\infty dzP(x-z,t)q(z)\\
P(x,0)=\delta(x)\end{array}\right..
\end{eqnarray}
For the characteristic function we therefore find
\begin{eqnarray}
\label{another_approach_indep_char}
\left\{\begin{array}{ll}\displaystyle\frac{\partial\hat{P}(k,t)}{\partial t}=
ikv\hat{P}(k,t)-r\hat{P}(k,t)+r\hat{P}(k,t)\hat{q}(k)\\
\hat{P}(k,0)=1\end{array}\right..
\end{eqnarray}
The solution of Eq.~\eqref{another_approach_indep_char} is
\begin{eqnarray}
\label{another_approach_indep_char_sol}
\hat{P}(k,t)=\exp(ikvt)\sum^\infty_{n=0}\frac{(rt)^n}{n!}\exp(-rt)(\hat{q}(k))^n,
\end{eqnarray}
which verifies our result \eqref{app_char_jump} for Poissonian resetting.

\section{Derivation of last resetting picture for dependent resetting amplitudes}
\label{dep_equal}

We now show the equivalence of the \emph{first resetting picture}
\begin{eqnarray}
\nonumber
&&P(x,t;x_0,t_0)=\Psi(t-t_0)G(x,t;x_0,t_0)+\int\limits_{t_0}^tdt_1\psi(t_1-t_0)
\int\limits_0^\infty\frac{dy}{y}G(y,t_1;x_0,t_0)\int\limits_0^ydx_1f_C\left(
\frac{x_1}{y}\right)P(x,t;x_1,t_1)\\
&&=\Psi(t-t_0)G(x,t;x_0,t_0)+\int\limits_{t_0}^t dt_1\psi(t_1-t_0)\int\limits_0
^\infty dyG(y,t_1;x_0,t_0)\int\limits_0^1dc_1f_C(c_1)P(x,t;c_1y,t_1),
\label{p_r_dep_first_app}
\end{eqnarray}
with $c_1=x_1/y$, and the \emph{last resetting picture}
\begin{eqnarray}
\nonumber
&&P(x,t;x_0,t_0)=\Psi(t-t_0)G(x,t;x_0,t_0)+\sum^\infty_{n=1}\int\limits_0^{t-t_0}
d\tau_n\int\limits_0^1dc_n\int\limits_0^\infty dy_n\\
\nonumber
&&\times\left(\prod^{n-1}_{i=1}\int\limits_0^{\tau_{n+1-i}} d\tau_{n-i}\psi(\tau
_{n+1-i}-\tau_{n-i})\int\limits_0^\infty dy_{n-i}G(y_{n+1-i},\tau_{n+1-i}+t_0;c
_{n-i}y_{n-i},\tau_{n-i}+t_0)\int\limits_0^1dc_{n-i}f_C(c_{n+1-i})\right)\\
&&\times f_C(c_1)\psi(\tau_1)G(y_1,\tau_1+t_0;c_0y_0,t_0)\Psi(t-t_0-\tau_n)G(x,t;
c_ny_n,\tau_n+t_0),
\label{p_r_dep_last_app}
\end{eqnarray}
with $c_0=1$ and $y_0=x_0$. Therefore,
\begin{eqnarray}
\nonumber
&&P(x,t;x',t')=\Psi(t-t')G(x,t;x',t')\\
&&+\sum^\infty_{n=1}\left(\prod^n_{j=1}\int\limits_{t_{j-1}}^t dt_j\psi(t_j
-t_{j-1})\int\limits_0^\infty dy_jG(y_j,t_j;c_{j-1}y_{j-1},t_{j-1})\int
\limits_0^1dc_j f_C(c_j)\right)\Psi(t-t_n)G(x,t;c_n y_n,t_n),
\label{p_r_dep_last_t_app}
\end{eqnarray}
with $t_0=t'$, $c_0=1$, and $y_0=x'$. The LHS of Eq.~\eqref{p_r_dep_first_app}
after substitution reads
\begin{eqnarray}
\nonumber
&&\mathrm{LHS}=P(x,t;x_0,t_0)=\Psi(t-t_0)G(x,t;x_0, t_0)\\
&&+\sum^\infty_{n=1}\left(\prod^n_{j=1}\int\limits_{t_{j-1}}^tdt_j\psi(t_j-
t_{j-1})\int\limits_0^\infty dy_jG(y_j,t_j;c_{j-1}y_{j-1},t_{j-1})\int
\limits_0^1dc_jf_C(c_j)\right)\Psi(t-t_n)G(x,t;c_n y_n,t_n).
\label{p_r_dep_last_t_LHS_app}
\end{eqnarray}
As $P(x,t;c_1y_1,t_1)$ in Eq.~\eqref{p_r_dep_first_app} has the initial value
$c_1y_1$ at $t_1$, these three variables have $1$ as lowest index, and we write
\begin{eqnarray}
\nonumber
&&P(x,t;x_1,t_1)=\Psi(t-t_1)G(x,t;c_1 y_1, t_1)\\
&&+\sum^\infty_{n=2}\left(\prod^n_{j=2}\int\limits_{t_{j-1}}^tdt_j\psi(t_j
-t_{j-1})\int\limits_0^\infty dy_jG(y_j,t_j;c_{j-1}y_{j-1},t_{j-1})\int
\limits_0^1dc_jf_C(c_j)\right)\Psi(t-t_n)G(x,t;c_n y_n,t_n).
\label{p_r_dep_last_app_1}
\end{eqnarray}
Substituting Eq.~\eqref{p_r_dep_last_app_1} into the RHS of
Eq.~\eqref{p_r_dep_first_app} we get
\begin{eqnarray}
\nonumber
&&\mathrm{RHS}=\Psi(t-t_0)G(x,t;x_0, t_0)\\
\nonumber
&&+\int\limits_{t_0}^tdt_1\psi(t_1-t_0)\int\limits_0^\infty dyG(y,t_1;x_0,t_0)
\int\limits_0^1dc_1f_C(c_1)\Psi(t-t_1)G(x,t;c_1y_1,t_1)\\
\nonumber
&&+\int\limits_{t_0}^tdt_1\int\limits_0^\infty dy\int\limits_0^1dc_1\sum^\infty
_{n=2}\left(\prod^{n}_{j=2}\int\limits_{t_{j-1}}^tdt_j\psi(t_j-t_{j-1})\int
\limits_0^\infty dy_jG(y_j,t_j;c_{j-1}y_{j-1},t_{j-1})\int\limits_0^1dc_jf_C(
c_j)\right)\\
\nonumber
&&\times\psi(t_1-t_0)G(y,t_1;x_0,t_0)f_C(c_1)\Psi(t-t_n)G(x,t;c_ny_n,t_n)\\
\nonumber
&&=\Psi(t-t_0)G(x,t;x_0, t_0)\\
\nonumber
&&+\int\limits_{t_0}^tdt_1\psi(t_1-t_0)\int\limits_0^\infty dy_1G(y_1,t_1;
x_0,t_0)\int\limits_0^1dc_1f_C(c_1)\Psi(t-t_1)G(x,t;c_1y_1,t_1)\\
&&+\sum^\infty_{n=2}\left(\prod^n_{j=1}\int\limits_{t_{j-1}}^tdt_j\psi(t_j
-t_{j-1})\int\limits_0^\infty dy_jG(y_j,t_j;c_{j-1}y_{j-1},t_{j-1})\int
\limits_0^1dc_jf_C(c_j)\right)\Psi(t-t_n)G(x,t;c_ny_n,t_n)
\end{eqnarray}
with $y_1=y$. Then,
\begin{eqnarray}
\nonumber
&&\mathrm{RHS}=\Psi(t-t_0)G(x,t;x_0, t_0)\\
&&+\sum^\infty_{n=1}\left(\prod^n_{j=1}\int\limits_{t_{j-1}}^tdt_j\psi(t_j
-t_{j-1})\int\limits_0^\infty dy_jG(y_j,t_j;c_{j-1}y_{j-1},t_{j-1})\int
\limits_0^1dc_jf_C(c_j)\right)\Psi(t-t_n)G(x,t;c_ny_n,t_n),
\label{p_r_dep_last_t_RHS_app}
\end{eqnarray}
and thus we have the identity $\mathrm{RHS}=\mathrm{LHS}$. Consequently
Eq.~\eqref{p_r_dep_last_t_app} solves the first resetting picture of
Eq.~\eqref{p_r_dep_first_app}. This implies that Eq.~\eqref{p_r_dep_last_t_app}
describes the RASR with a dependent resetting amplitude. If we show that
Eq.~\eqref{p_r_dep_last_t_app} and the last resetting picture of
Eq.~\eqref{p_r_dep_last_app} are equal, this means that both mathematical
representations are equivalent. To proceed,
\begin{eqnarray}
\nonumber
&&\mathrm{LHS}=P(x,t;x_0,t_0)=\Psi(t-t_0)G(x,t;x_0,t_0)\\
\nonumber
&&+\sum^\infty_{n=1}\left(\prod^n_{j=1}\int\limits_{t_{j-1}}^tdt_j\psi(t_{j}
-t_{j-1})\int\limits_0^\infty dy_jG(y_j,t_j;c_{j-1}y_{j-1},t_{j-1})\int
\limits_0^1dc_jf_C(c_j)\right)\Psi(t-t_n)G(x,t;c_ny_n,t_n)\\
\nonumber
&&=\Psi(t-t_0)G(x,t;x_0,t_0)\\
\nonumber
&&+\sum^\infty_{n=1}\left(\prod^n_{j=1}\int\limits_{\tau_{j-1}}^{t-t_0}d\tau_j
\psi(\tau_j-\tau_{j-1})\int\limits_0^\infty dy_jG(y_j,\tau_j+t_0;c_{j-1}y
_{j-1},\tau_{j-1}+t_0)\int\limits_0^1dc_jf_C(c_j)\right)\\
&&\times\Psi(t-\tau_n-t_0)G(x,t;c_n y_n,\tau_n+t_0),
\end{eqnarray}
with $\tau_j=t_j-t_0$ for $1\le j\le n$. If we now use
Eq.~\eqref{identity} with the substitutions
\begin{eqnarray}
\label{dep_subs}
\left\{\begin{array}{l}
\eta_1(t_j,t_{j-1},y_j,y_{j-1},z_j,z_{j-1})=\psi(t_j-t_{j-1})G(y_j,t_j;z_{j-1}
y_{j-1},t_{j-1})f_C(z_j)\\[0.2cm]
\eta_2(x,t,t_n,y_n,z_n)=\Psi(t-t_0)G(x,t;z_n y_n,t_n)\\[0.2cm]
t_j=\tau_j+t_0, \,\,\, z_j=c_j\\[0.2cm]
A_1=0,\,\,\, A_2=\infty,\,\,\, A_3=0,\,\,\, A_4=1\end{array}\right.
\end{eqnarray}
for $1\le j\le n$, we get
\begin{eqnarray}
\nonumber
&&\mathrm{LHS}=\Psi(t-t_0)G(x,t;x_0,t_0)+\sum^\infty_{n=1}\int\limits_0^{t-t_0}
d\tau_n\int\limits_0^1dc_n\int\limits_0^\infty dy_n\left(\prod^{n-1}_{i=1}\int
\limits_0^{\tau_{n+1-i}}d\tau_{n-i}\psi(\tau_{n+1-i}-\tau_{n-i})\right)\\
\nonumber
&&\times\left(\prod^{n-1}_{i=1}\int\limits_0^\infty dy_{n-i}G(y_{n+1-i},
\tau_{n+1-i}+t_0;c_{n-i}y_{n-i},\tau_{n-i}+t_0)\int\limits_0^1dc_{n-i}
f_C(c_{n+1-i})\right)\\
&&\times f_C(c_1)\psi(\tau_1)G(y_1,\tau_1+t_0;x_0,t_0)\Psi(t-t_0-\tau_n)
G(x,t;c_n y_n,\tau_n +t_0),\label{dep_last_derive}
\end{eqnarray}
which is exactly the last resetting picture.

If we assume that the free propagator is homogeneous in space and in time, 
the stochastic process will be also homogeneous in time but not in space,
$G(x,t;x_0,t_0)=G(x-x_0,t-t_0;0,0)\Rightarrow P(x,t;x_0,t_0)=P(x,t-t_0;x_0,0)$.
By assuming $G(x,t;x_0,t_0)=G(x-x_0,t-t_0;0,0)$ the density $P(x,t;x_0,t_0)$,
Eq.~\eqref{dep_last_derive}, becomes
\begin{eqnarray}
\nonumber
&&P(x,t;x_0,t_0)=\Psi(t-t_0)G(x-x_0,t-t_0;0,0)+\sum^\infty_{n=1}\int\limits_0^{t-t_0}
d\tau_n\int\limits_0^1dc_n\int\limits_0^\infty dy_n\left(\prod^{n-1}_{i=1}\int
\limits_0^{\tau_{n+1-i}}d\tau_{n-i}\psi(\tau_{n+1-i}-\tau_{n-i})\right)\\
\nonumber
&&\times\left(\prod^{n-1}_{i=1}\int\limits_0^\infty dy_{n-i}G(y_{n+1-i}-c_{n-i}y_{n-i},
\tau_{n+1-i}-\tau_{n-i};0,0)\int\limits_0^1dc_{n-i} f_C(c_{n+1-i})\right)\\
&&\times f_C(c_1)\psi(\tau_1)G(y_1-x_0,\tau_1; 0,0)\Psi(t-t_0-\tau_n)
G(x- c_n y_n,t-t_0-\tau_n;0,0).\label{dep_in_homogen_app}
\end{eqnarray}
On the right hand side of Eq.~\eqref{dep_in_homogen_app} $t$ and $t_0$ only arise
as differences $t-t_0$, however $x$ and $x_0$ occur as single term. Thus, $G(x,t;
x_0,t_0)=G(x-x_0,t-t_0;0,0)\Rightarrow P(x,t;x_0,t_0)=P(x,t-t_0;x_0,0)\neq P(x-x_0
,t-t_0;0,0)$, which proves our claim.

\section{First and second derivatives of Eq.~(\ref{app_dep_bal_lap_lap})
with respect to the Laplace variable $u$}
\label{appe}

The first derivative of Eq.~\eqref{app_dep_bal_lap_lap} reads
\begin{eqnarray}
\nonumber
\tilde{\bar{P}}'(u,s;x_0)&=&\sum^\infty_{n=0}\tilde{\Psi}(s+uv)\left(\prod^n_{k=1}
\int\limits_0^1dc_kf_C(c_k)\tilde{\psi}\left(s+uv\prod^k_{i=1}c_i\right)\right)
\exp\left(-ux_0\prod^n_{j=0}c_j\right)\\
&&\times\left(\frac{v\tilde{\Psi}'(s+uv)}{\tilde{\Psi}(s+uv)}+v\sum_{l=1}^n\frac{
\tilde{\psi}'\left(s+uv\prod^l_{i=1}c_i\right)\prod^l_{i=1}c_i}{\tilde{\psi}
\left(s+uv\prod^l_{i=1}c_i\right)}-x_0\prod^n_{j=0}c_j\right).
\label{app_dep_bal_lap_lap_first_der}
\end{eqnarray}
Using Eq.~\eqref{app_dep_bal_lap_lap_first_der} and with the notation $\langle c
\rangle=\int^1_0cf_C(c)dc$ this expression is rewritten as
\begin{eqnarray}
\nonumber
\tilde{\bar{P}}'(0,s;x_0)&=&\sum^\infty_{n=0}\tilde{\Psi}(s)\tilde{\psi}^n(s)\left(
v\frac{\tilde{\Psi}'(s)}{\tilde{\Psi}(s)}+v\frac{\tilde{\psi}'(s)}{\tilde{\psi}
(s)}\sum_{l=1}^n\langle c\rangle^l-x_0\langle c\rangle^n\right)\\
&=&\sum^\infty_{n=0}\left(v\tilde{\psi}^n(s)\tilde{\Psi}'(s)+v\tilde{\psi}^{n-1}
(s)\tilde{\psi}'(s)\tilde{\Psi}(s)\frac{\langle c\rangle-\langle c\rangle^{n+1}}
{1-\langle c\rangle}-x_0\tilde{\psi}^n(s)\tilde{\Psi}(s)\langle c \rangle^n\right).
\label{app_dep_bal_lap_lap_first_der_zero}
\end{eqnarray}

The second derivative of Eq.~\eqref{app_dep_bal_lap_lap} is
\begin{eqnarray}
\nonumber
\tilde{\bar{P}}''(u,s;x_0)&=&\sum^\infty_{n=0}\tilde{\Psi}(s+uv)\left(\prod^n
_{k=1}\int\limits_0^1dc_kf_C(c_k)\tilde{\psi}\left(s+uv \prod^{k}_{i=1}c_i\right)
\right)\exp\left(-ux_0\prod^n_{j=0}c_j\right)\\
\nonumber
&&\times\left(\left(\frac{v\tilde{\Psi}'(s+uv)}{\tilde{\Psi}(s+uv)}+v\sum_{l=1}^n 
\frac{\tilde{\psi}'\left(s+uv\prod^l_{i=1}c_i\right)\prod^l_{i=1}c_i}{\tilde{\psi}
\left(s+uv\prod^l_{i=1}c_i\right)}-x_0\prod^n_{j=0}c_j\right)^2\right.\\
\nonumber
&&\left.+v^2\frac{\tilde{\Psi}''(s+uv)\tilde{\Psi}(s+uv)-\left(\tilde{\Psi}'
(s+uv)\right)^2}{\tilde{\Psi}^2(s+uv)}\right)\\
\nonumber
&&+\sum^\infty_{n=0}\tilde{\Psi}(s+uv)\left(\prod^n_{k=1}\int\limits_0^1dc_kf_C
(c_k)\tilde{\psi}\left(s+uv\prod^k_{i=1}c_i\right)\right)\exp\left(-ux_0\prod^n
_{j=0}c_j\right)\\
&&\times\left(v^2\sum_{l=1}^n\frac{\tilde{\psi}\left(s+uv\prod^l_{i=1}c_i\right) 
\tilde{\psi}''\left(s+uv\prod^l_{i=1}c_i\right)-\left(\tilde{\psi}'\left(s+uv
\prod^l_{i=1}c_i\right)\right)^2}{\tilde{\psi}^2\left(s+uv\prod^l_{i=1}c_i\right)}
\prod^l_{i=1}c^2_i\right).
\label{app_dep_bal_lap_lap_second_der}
\end{eqnarray}
With the definition $\langle c^2\rangle=\int^1_0c^2f_C(c)dc$ we further transform
this expression to
\begin{eqnarray}
\nonumber
&&\tilde{\bar{P}}''(0,s;x_0)
=\sum^\infty_{n=0}\tilde{\Psi}(s)\tilde{\psi}^n(s)\left(v^2\frac{\tilde{\Psi}
''(s)\tilde{\Psi}(s)-\left(\tilde{\Psi}'(s)\right)^2}{\tilde{\Psi}^2(s)}+v^2\frac{
\tilde{\psi}''(s)\tilde{\psi}(s)-\left(\tilde{\psi}'(s)\right)^2}{\tilde{\psi}^2
(s)}\sum_{l=1}^n\langle c^2\rangle^l+2v^2\frac{\tilde{\Psi}'(s)\tilde{\psi}'(s)}{
\tilde{\Psi}(s)\tilde{\psi}'(s)}\sum_{l=1}^n\langle c\rangle^l\right)\\
\nonumber
&&+\sum^\infty_{n=0}\tilde{\Psi}(s)\tilde{\psi}^n(s)\left(v^2\frac{\left(\tilde{
\Psi}'(s)\right)^2}{\tilde{\Psi}^2(s)}+v^2\frac{\left(\tilde{\psi}'(s)\right)^2}{
\tilde{\psi}^2(s)}\sum_{l=1}^n\left(\langle c^2\rangle^l+2\sum_{m=1}^{l-1}\langle
c^2\rangle^m\langle c\rangle^{l-m}\right)+x^2_0\langle c^2\rangle^n\right)\\
\nonumber
&&-\sum^\infty_{n=0}\tilde{\Psi}(s)\tilde{\psi}^n(s)\left(2vx_0\frac{\tilde{\Psi}
'(s)}{\tilde{\Psi}(s)}\langle c\rangle^n+2vx_0\frac{\tilde{\psi}'(s)}{\tilde{
\psi}(s)}\sum_{l=1}^n\langle c^2\rangle^l\langle c\rangle^{n-l}\right).
\end{eqnarray}

$\tilde{\bar{P}}''(0,s;x_0)$ can now be simplified to
\begin{eqnarray}
\nonumber
&&\tilde{\bar{P}}''(0,s;x_0)=\sum^\infty_{n=0}v^2\left(\tilde{\psi}^n(s)\tilde{
\Psi}''(s)+\tilde{\psi}^{n-1}(s)\tilde{\psi}''(s)\tilde{\Psi}(s)\frac{\langle
c^2\rangle-\langle c^2\rangle^{n+1}}{1-\langle c^2\rangle}+2\tilde{\psi}^{n-1}(s)
\tilde{\psi}'(s)\tilde{\Psi}'(s)\frac{\langle c\rangle-\langle c\rangle^{n+1}}{1-
\langle c\rangle}\right)\\
\nonumber
&&+\sum^\infty_{n=0}2v^2\tilde{\Psi}(s)(\tilde{\psi}')^2 (s)\left(\frac{\tilde
\psi^{n-2}(s)\langle c^2\rangle\langle c\rangle}{(1-\langle c^2\rangle)(1-
\langle c\rangle)}+\frac{\tilde{\psi}^{n-2}(s)\langle c^2\rangle\langle c\rangle^{
n+1}}{(\langle c\rangle-\langle c^2\rangle)(\langle c\rangle-1)}+\frac{\tilde{\psi}
^{n-2}(s)\langle c\rangle\langle c^2\rangle^{n+1}}{(\langle c\rangle-\langle c^2
\rangle)(1-\langle c^2\rangle)}\right)\\
&&-\sum^\infty_{n=0} 2vx_0\left(\tilde{\psi}^n(s)\tilde{\Psi}'(s)\langle c \rangle^n
+\tilde{\psi}^{n-1}(s)\tilde{\psi}'(s)\tilde{\Psi}(s)\frac{\langle c^2\rangle^{n+1}-
\langle c\rangle^n\langle c^2\rangle}{\langle c^2\rangle-\langle c\rangle}\right)
+\sum^\infty_{n=0}x^2_0\tilde{\psi}^n(s)\tilde{\Psi}(s)\langle c^2\rangle^n.
\label{app_dep_bal_lap_lap_second_der_zero}
\end{eqnarray}

\twocolumngrid

\end{document}